\documentclass[pra,aps,showpacs,twocolumn,superscriptaddress]{revtex4}
\usepackage{graphicx,amsfonts,amsmath,amssymb,xcolor}
\usepackage{hyperref}
\hypersetup{colorlinks=true,citecolor=blue}
\graphicspath{{images/}}
\begin{document}
\title{Resonator neuron and triggering multipulse excitability in laser with injected signal}
\author{A. Dolcemascolo}
\affiliation{Universit$\acute{e}$ C\^ote d'Azur, CNRS UMR 7335, Institut de physique de Nice, 1361 Route des Lucioles, F-06560 Valbonne, France}
\author{B. Garbin}
\affiliation{The Dodd-Walls Centre for Photonic and Quantum Technologies, Department of Physics, The University of Auckland, Auckland, New Zealand}
\author{B. Peyce}
\affiliation{Universit$\acute{e}$ C\^ote d'Azur, CNRS UMR 7335, Institut de physique de Nice, 1361 Route des Lucioles, F-06560 Valbonne, France}
\author{R. Veltz}
\affiliation{Inria Sophia Antipolis, MathNeuro Team, 2004 route des Lucioles - BP93, 06902 Sophia Antipolis, France}
\author{S. Barland}
\affiliation{Universit$\acute{e}$ C\^ote d'Azur, CNRS UMR 7335, Institut de physique de Nice, 1361 Route des Lucioles, F-06560 Valbonne, France}
\date{\today}%

\begin{abstract}

	Semiconductor lasers with coherent forcing are expected to behave similarly to simple neuron models in response to external perturbations, as long as the physics describing them can be approximated by that of an overdamped pendulum with fluid torque. Beyond the validity range of this approximation, more complex features can be expected. We perform experiments and numerical simulations which show that the system can display resonator and integrator features depending on parameters and that multiple pulses can be emitted in response to larger perturbations.

\end{abstract}

\pacs{42.55.Px,42.65.Sf,05.45.Xt}

\maketitle


\section{Introduction}

The quest for new approaches to computing takes many forms and one of the most exciting is certainely the use of dynamical systems, in particular with the design of brain-inspired processors \cite{poon2011neuromorphic}. Most of these approaches are based on electronic platforms, which are the most natural and immediate choice as most computing devices relie on electronics already. However, since the transport of information at large distance and increasingly also at short distance is based on light, there is an interest in offloading part of the data processing to optical devices which would naturally interface with the optical layer. For instance, photonic reservoir computers (see \textit{eg} \cite{larger2012photonic,brunner2013parallel,vandoorne2014experimental,duport2016fully,larger2017high}), nanophotonic circuits \cite{shen2017deep} or even a multiple scattering method \cite{saade2016random} aim at leveraging complex dynamics in optical systems to provide part or all of the computation stages required to accomplish computing tasks, even when the components of the system do not attempt to emulate neurons. A complementary approach consists in achieving with optical devices activation functions which actually mimick that of biological neurons, an approach sometimes termed photonic spike processing \cite{prucnal2016recent}. 

Along this last line, one of the landmarks of neurosciences is the analysis of the electrical response of a neural cell to external perturbations \cite{hodgkin1952measurement,hodgkin1952quantitative}. The all-or-nothing response of the cell, which is triggered only for perturbations which are large enough but does not depend on the perturbation itself once the threshold is overcome, is widely considered as a key ingredient for the processing of information by neural cells. For this reason, this type of "excitable" response has been investigated in many physical systems and in particular in optical devices. In this specific context, several possible dynamical scenarii have been analyzed: close to a saddle-node bifurcation \cite{coullet1998optical,giudici1997andronov,goulding2007excitability}, weakly saturated Hopf bifurcation \cite{barland2003experimental} and saddle-loop bifurcation \cite{dubbeldam1999excitability,larotonda2002experimental}. Most recent approaches in this direction are based on potentially integrable components such as semiconductor lasers \cite{turconi2013control,garbin2013control,sorrentino2015effects} sometimes with polarization effects \cite{hurtado2010optical,hurtado2012investigation,hurtado2015controllable,deng2017controlled}, silicon microrings \cite{van2012cascadable}, micropillars with integrated saturable absorber \cite{barbay2011excitability,selmi2014relative,selmi2015temporal} and resonant tunneling diodes \cite{romeira2013excitability,romeira2016regenerative}. 

Interestingly each of these systems differ not only by their physical nature but also by the dynamical mechanisms which are at the origin of their excitable character. This is important since, depending on the type of bifurcation which causes the excitable response, neurons can have different additional properties with respect to repeated perturbations \cite{izhipaper}, which of course strongly influences the dynamics of coupled systems and in turn their computational properties. In particular some neurons have the capability to \textit{integrate} several sub-threshold repeated perturbations, leading to an excitable response when a sufficient number of perturbations are applied repeatedly. In optics, this behavior typical of "integrator" neurons has been observed in \cite{selmi2015temporal} and is also expected to exist in the case of a laser with injected signal when the dynamics can be reduced to that of the optical phase, \textit{i.e.} when the excitable response consists only of a $2\pi$ rotation of the laser phase with respect to the injection signal. Other neurons though, have the specific property of responding to repeated subthreshold perturbation only if these are adequately separated in time. In the case of optics, this behavior has not been observed yet, but the laser with injected signal is certainely a good candidate for this kind of observation as soon as the dynamics can not be completely reduced to that of the optical phase, such as when multipulse excitability \cite{wieczorek2002multipulse,kelleher2011excitability} is present.  In the following, we analyze the response of a laser with injected signal close to unlocking transition, where the control of excitable pulses and the existence of a refractory time were demonstrated recently \cite{turconi2013control,garbin2013control}. We demonstrate that indeed two perturbations can be integrated by the system and lead to an excitable response even when each of them would not be sufficient to trigger a spike. At variance with a pure integrate and fire neuron though, we show that there is an optimum value for the time separation between these two perturbations, for which their efficiency is maximum. We analyse these results from two perspectives. First we show that an \textit{ad hoc} generalization of the Adler equation \cite{adler1946study} is sufficient to reproduce the results and second we study this same behavior in a realistic model for a Class-B laser with optical injection.

\section{Experimental setup}
\begin{figure}
\includegraphics[width=0.48\textwidth]{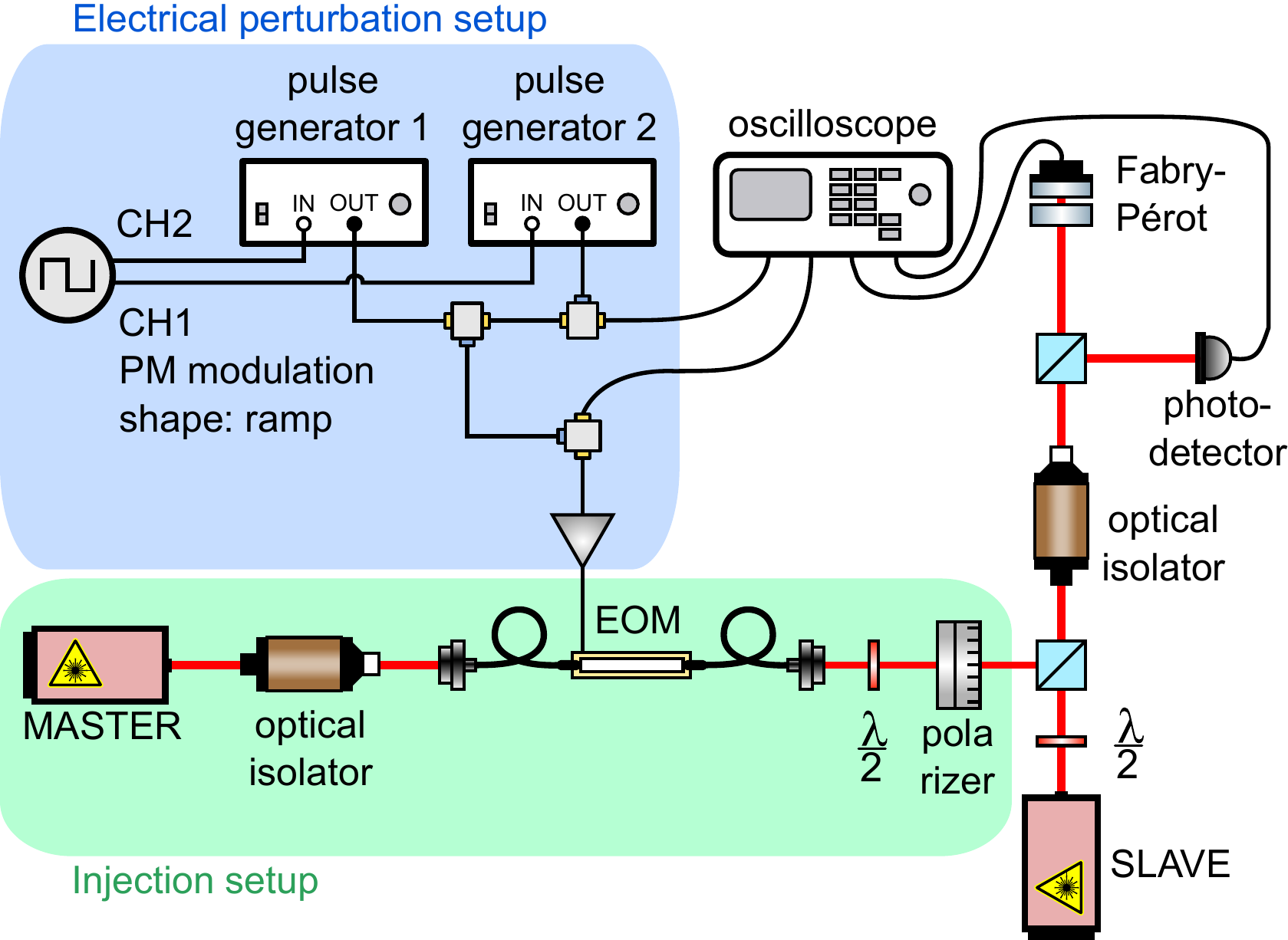}%

\caption{Schematic of the experimental setup. The injection setup is highlighted in green, and the electrical perturbation setup is highlighted in blue.}
\label{fig_setup}
\end{figure}

The experimental setup is that of a VCSEL (Vertical Cavity Surface Emitting Laser) with optical injection, as showed in Fig.~\ref{fig_setup} . It is exactly the same setup used in \cite{bruno13}, with the addition of a more involved electrical perturbation setup. 

The aim is to inject the signal coming from a \textit{master} laser  into the \textit{slave} laser (a VCSEL). In FIG. \ref{fig_setup} we can see the injection setup, which is composed of: the \textit{master} laser (tunable via an external grating); an optical isolator to prevent unwanted reflections from reaching the laser back; a fiber-coupled electro-optic modulator (EOM) and a half wave plate plus a polarizer (with vertical orientation) to modulate the intensity of the injected beam. The EOM allows us to apply a phase perturbation to the master signal with shape and amplitude that is determined by a voltage input.

After that, the master signal is injected into the slave laser through a beam splitter. A half wave plate is placed just at the output of the collimating lens of the slave laser in order to adjust the polarization of the slave laser to the vertical axis. The output from the slave laser is then first conveyed through an optical isolator to prevent again spurious reflections towards the slave laser, and then sent to a 9-GHz photodetector and a Fabry-Pérot interferometer for spectral monitoring. The detection signals are acquired with a 12.5-GHz bandwidth real-time oscilloscope.

\begin{figure}[b]
\centering
\includegraphics[width=0.48\textwidth]{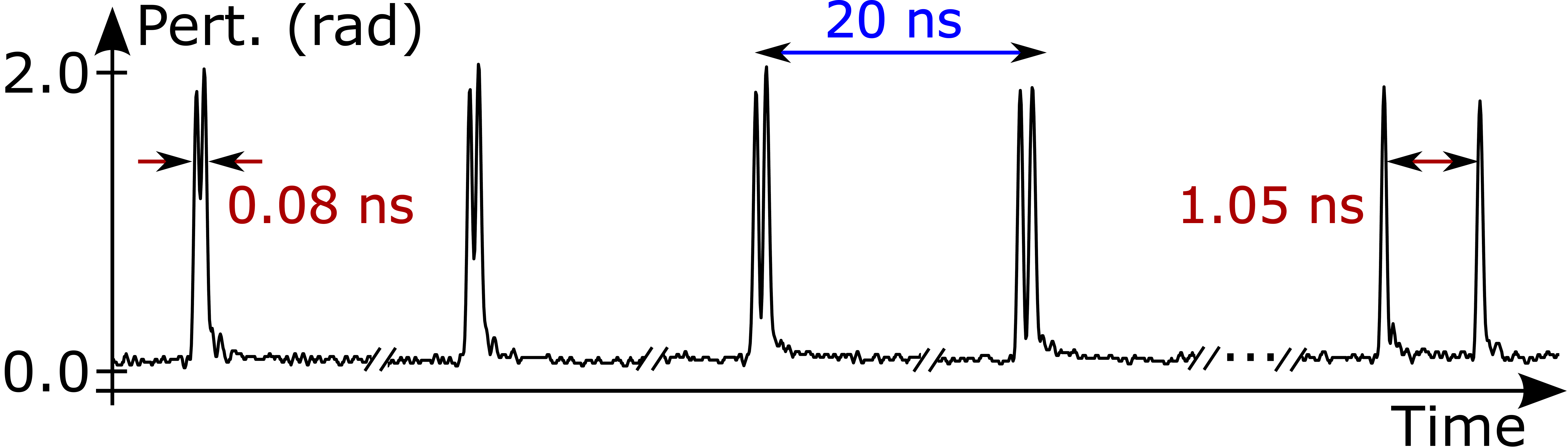}%
\caption{\label{fig_perturbazione} Simplified depiction of the shape of the periodic electrical perturbation sent to the EOM. The delay times between each couple of pulses increases gradually from a minimum of 0.08 ns to a maximum of 1.05 ns. Each couple is separated by its neighbours by a 20 ns delay. The same perturbation is repeated periodically after completion, where each period takes 2.5 $\mu$s in total. }
\end{figure}

The application of repeated perturbations in the phase of the injection beam is obtained via the application of repeated voltage perturbations to the EOM. The applied perturbation has the shape of FIG. \ref{fig_perturbazione}. It consists of a series of pulses, always in pair, with a constant height of 7.4 V (2 radians) and a duration of about 0.12 ns, where the delay between the first pulse and second pulse can be increased from a minimum of 0.08 ns to a maximum of 1.05 ns (the delay is defined as the time between the two maxima). The experiment is performed in a repetitive way to allow for statistics. In order to consider the many realizations of the experiment as independent, they must be sufficiently separated in time. In this case, the time between two realization is 20 ns, which is very long as compared to the previously observed refractory time, of the order of 1~ns \cite{garbin2017refractory}. In order to scan the delay in an automated way, we continuously scan the delay between the two perturbations, from 0.08 to 1.05~ns. This whole measurement is then repeated every 2.5 $\mu$s, starting again from the minimum delay all the way to the maximum delay. We require this kind of perturbation in order to explore all of the different delays in one go, without letting too much time pass between one delay and the next, so that we can assume that the parameters of the system are stationary during the acquisitions.

To obtain this kind of perturbation we have assembled the electrical perturbation setup as in FIG. \ref{fig_setup}. We use two pulse generators: an Alnair Labs EPG-200B-0050-0250 (first pulse generator) and an Alnair Labs EPG-210B-0050-S-P-T-A (second pulse generator). They respond to an input raising front by generating a pulse with constant amplitude and tunable width. Each of them is then able to generate a 50~ps duration pulse. In order to progressively increase the delay between the creation of the two pulses in a pair, we trigger the first pulse generator by a 50 MHz square wave, and the second pulse generator by a second 50 MHz square wave with 800 kHz phase modulation in the shape of a down-ramp. Since the two square waves are synchronized (because they are generated by the same signal generator) this creates a periodic shift between the two square waves, that later translates into a delay between the creation of the two pulses by the pulse generators. The pulses are later added by an RF combiner, and amplified up to 7.5 V before entering the EOM. We also detect the signal coming from the second pulse generator (used as a trigger for the oscilloscope), and the signal going to the EOM, in order to record at the same time the perturbation and the response of the system.

\section{Results}
\label{sec:experiment}

\begin{figure}[h]
\includegraphics[width=0.42\textwidth]{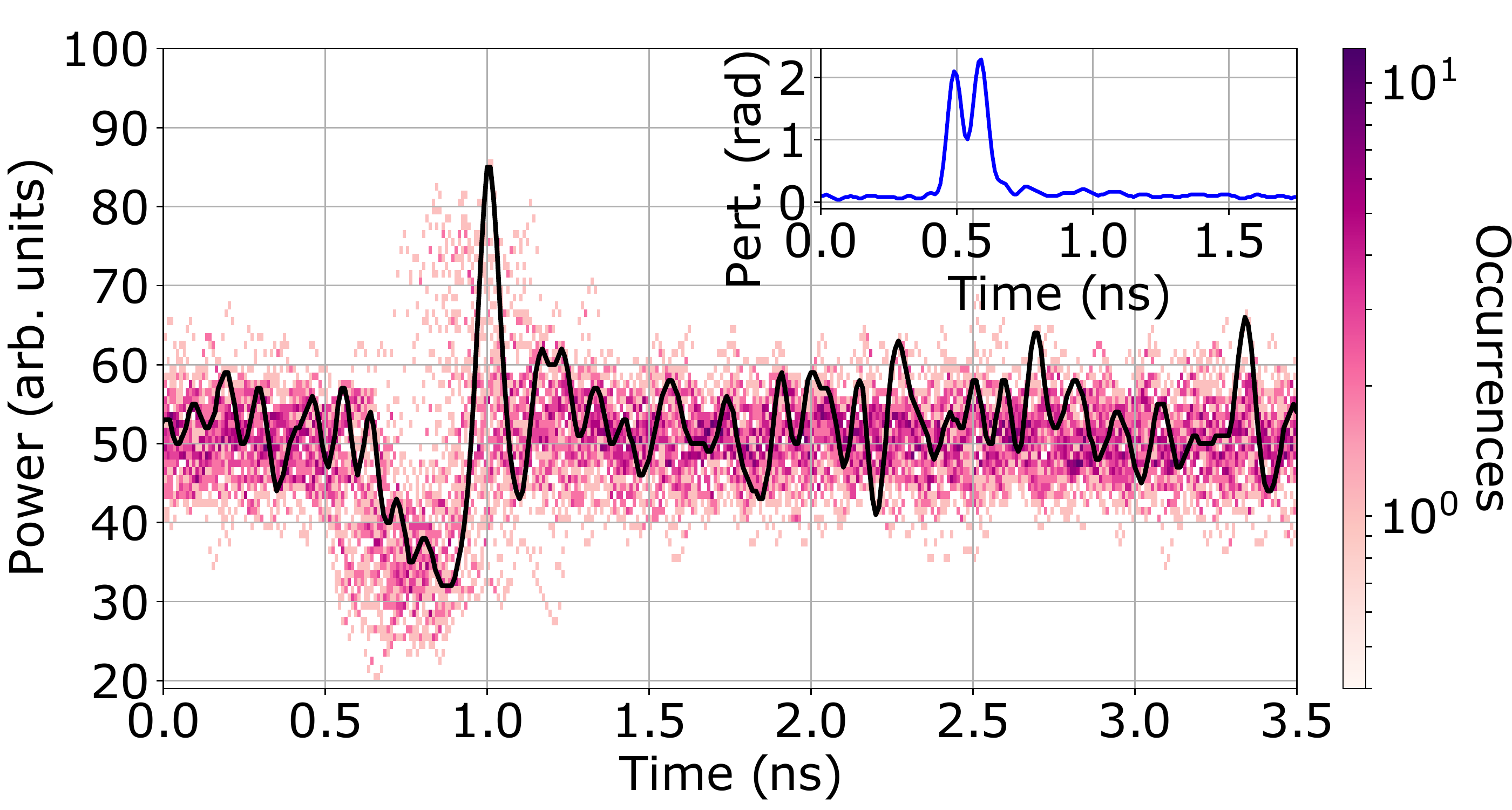}
\includegraphics[width=0.42\textwidth]{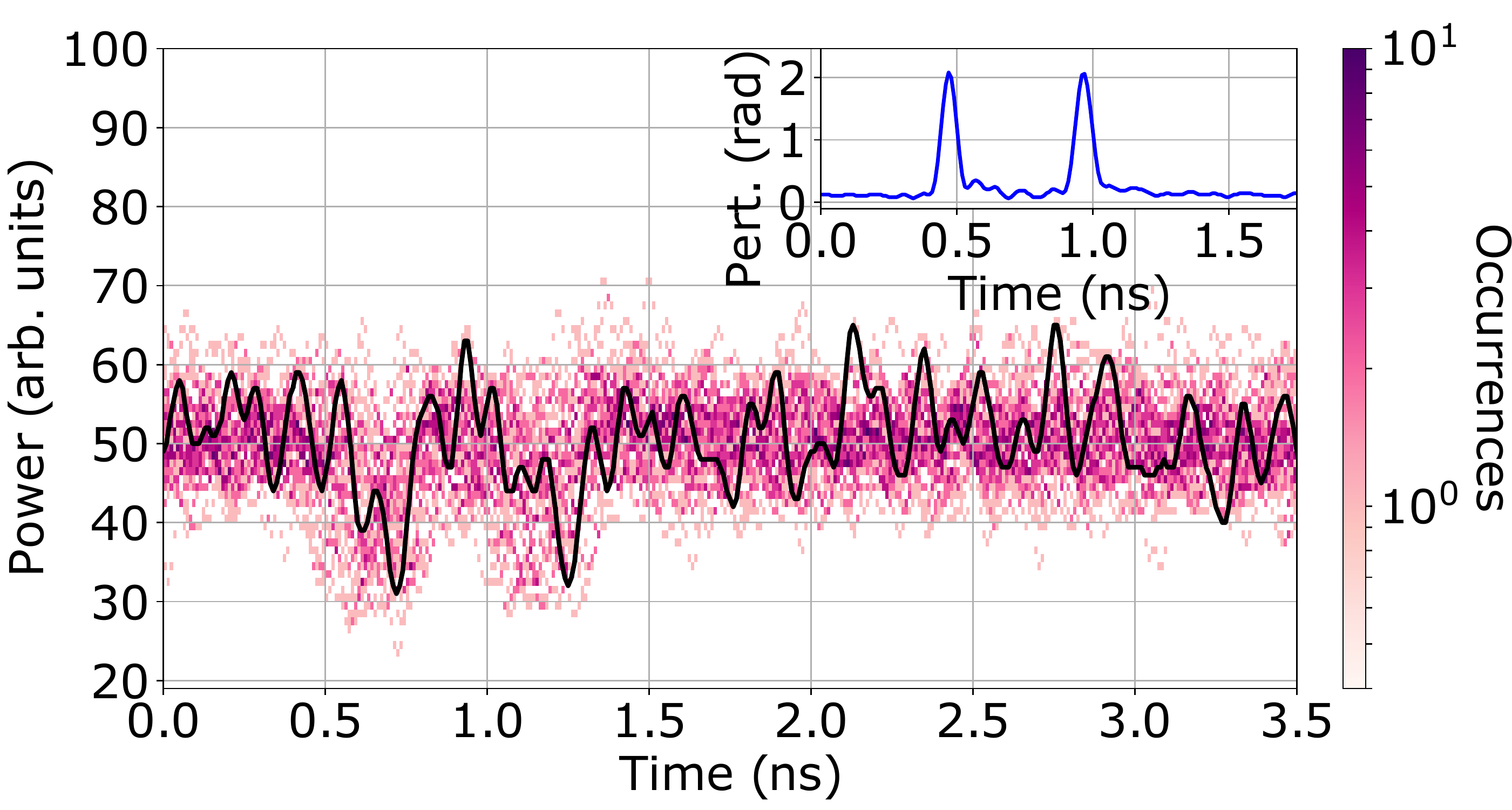}%
	\caption{\label{timetraces_exp} Experimentally measured time traces of the response of the system when the two perturbations are separated by 0.10~ns (top) and 0.50~ns (bottom). The emitted power (DC-level) is about 500$\mu$W and the injected power is 4.6$\mu$W. Insets: shape of the perturbations. Pumping current: 1.023~mA. 40 realizations are superimposed and show that on the bottom trace, no excitable pulse was observed. One example realization is shown as the black trace.}
\end{figure}

In this study we want to probe the integration property of the optical device. After having it prepared in an excitable regime by placing it inside the locking region close to the unlocking boundary defined by the saddle-node bifurcation (procedure already described in \cite{garbin2017refractory}), we then apply a series of perturbations which are, by themselves, under threshold. The integration behaviour of the system would be revealed if, given two or more under-threshold perturbations that are close in time, we were nevertheless able to observe an excitable response. The type of perturbations that we applied consists of a series of couples of pulses with different delays between them, as described before. Two examples for two different delays are shown in Fig. \ref{timetraces_exp}, where we show the persistence histogram for the two delays of 0.10 ns and 0.50 ns. When the two perturbations are sufficiently close in time (0.10~ns) an excitable spike can be generated, while no spike is generated when the two perturbations are too separated in time (0.50 ns). 

\begin{figure}[t]
\includegraphics[width=0.47\textwidth]{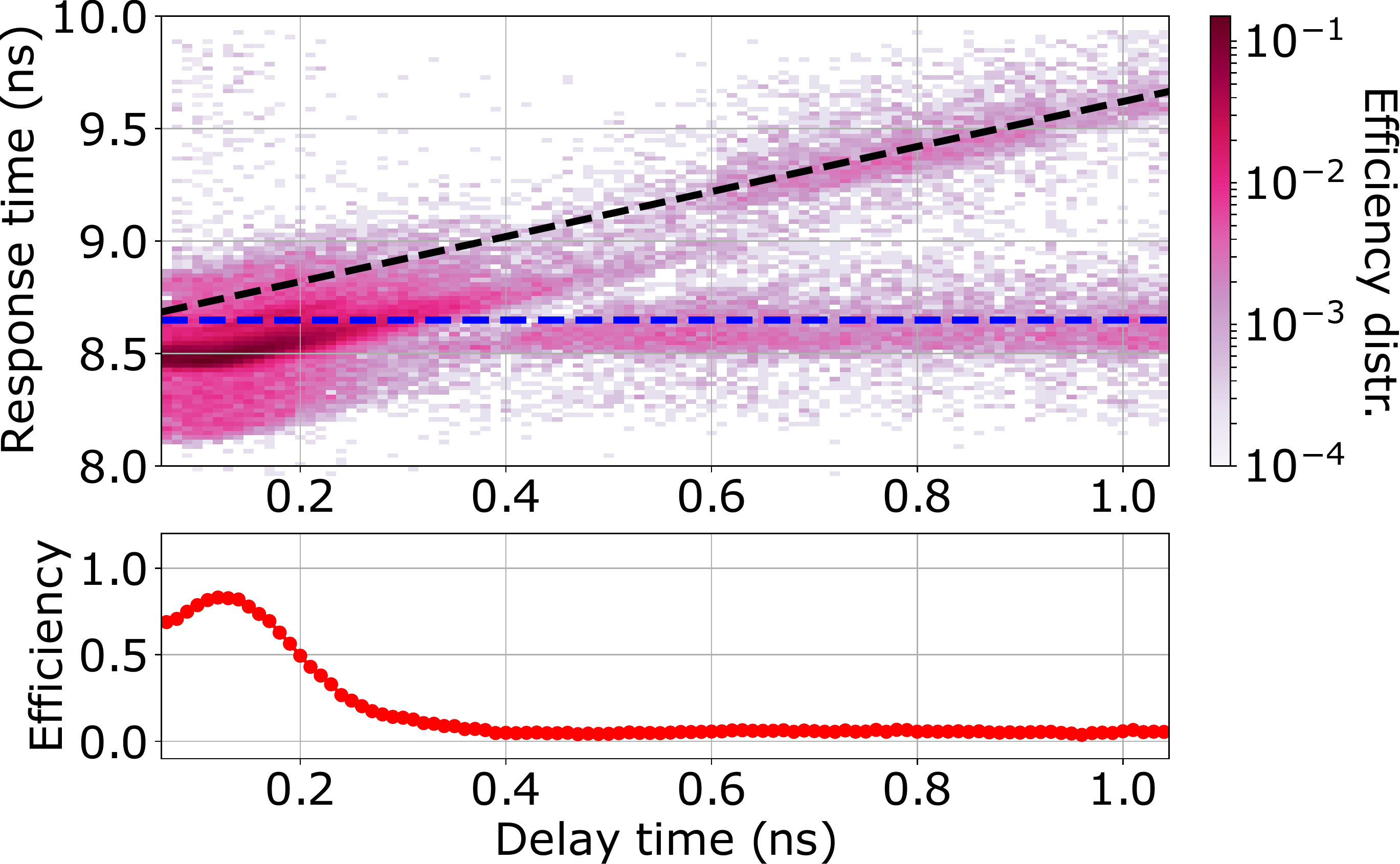}%
\caption{\label{fig_histo} Experimental response time histogram and efficiency curve of the perturbation for around 4000 events for each delay. ($S=1.023\, A$). The histogram is normalized so that each vertical slice for a single delay sums up to the corresponding efficiency value.}
\end{figure}

To quantify our results, we calculate the efficiency of each pair of perturbations, where the efficiency is defined as the number of excitable responses over the number of perturbations applied. In our analysis, we define an excitable response as a pulse whose amplitude is bigger then a certain threshold that we define \textit{a posteriori} (in this case, that is bigger than 23 arbitrary units from the baseline of the intensity signal). The results are shown on the bottom panel of Fig.~\ref{fig_histo}. We observe that the efficiency curve does increase for small delays, but it also presents a maximum at around 0.12 ns. This can be interpreted as a \textit{resonant feature}, that is the system has a higher probability of generating a pulse if we perturb it twice with the correct temporal separation. Notably, this optimal temporal separation is very similar to the temporal separation between subsequent spikes in the case of multipulse emission discussed in section~\ref{sec:multipulse}.

We have also analysed the response time of the excitable response for different delays, as shown on top panel of Fig.~\ref{fig_histo}. Note that the absolute value of the response time shown here is defined as the time difference between the maximum of the excitable pulse and the trigger time of the data acquisition system. Thus, it includes a very large offset which is purely of instrumental origin and not related to laser physics. The histogram is normalized to the efficiency curve, so that each vertical slice for a single delay sums up to the corresponding efficiency value. For high delays, the dashed blue line (horizontal) is the average arrival time of an excitable response generated by the first pulse of the perturbation (which is constant), while the black dashed line is the average arrival time of a response created by the second pulse (which moves away as the delay increases). We underline that the phenomenon of a single perturbation triggering a response is very rare, as indicated by the very low efficiency (bottom panel) for large delays. It is however revealed by the use of the logarithmic color scale for histograms or arrival times shown on the top panel. We extend these lines for smaller delays in order to have a frame of reference. For small delays, we observe that the responses happen mostly in a narrow range, with a big spread in time. The core with higher probability of arrival times also happens to be below the blue dashed line. This means that the excitable responses generated by resonance of the two pulses are created a bit faster then the response of a single perturbation which is coherent with the observation that a stronger perturbation can generate a response faster than a weaker perturbation \cite{garbin2017refractory}. Another observation is that, even though it is not visible in the efficiency curve, there is still some interaction between the two responses for delays between 0.4 and 0.6 ns. In this range the response time histogram shows gaps and lines that are not coherent with a single sum of the two perturbations as with longer delays. This weak interaction slowly disappears for delays longer than 0.7-0.8 ns, which is the same order of magnitude of the interaction time between two perturbations already observed in \citep{garbin2017refractory}.

\section{Ad hoc modelling: beyond the overdamped pendulum}

Interestingly, the reduction of the dynamics of a laser with injected signal to that of the optical phase leads to describing the laser with the Adler equation, which also describes an overdamped mechanical oscillator with forcing \cite{coullet2005damped} and is also known in neurosciences as the $\theta$-model or Ermentrout-Kopell canonical model \cite{ermentrout1986parabolic,izhipaper}.
 In \cite{coullet2005damped}, the case of finite damping (presence of inertia) was also analyzed, leading to bistability between the locked and the oscillating solution.  As a pure \textit{ad hoc} phenomenological modelling, we consider the response of a damped (but not overdamped) pendulum with fluid torque, \textit{ie} an Adler equation with a small inertial term. In practice, what we model here is that, after the first pulse perturbation, the pendulum does not simply relax into the fixed point but instead oscillates around it for a few times. If we can time the second perturbation so that it kicks the pendulum when it oscillates closer to the unstable point, then it is more likely that an excitable response will be triggered. This added dimension (inertia) in the phase space has been shown to heavily impact the interspike time distribution in the case of an excitable system with noise \cite{eguia2000distribution} and also to strongly impact the transition to synchrony in a modified Kuramoto model \cite{olmi2014hysteretic}. Here we check numerically that indeed, this new dimension adds to the integrate and fire mechanism and leads to a maximum in the efficiency curves as observed experimentally. 

\begin{figure}[t]
\includegraphics[width=0.47\textwidth]{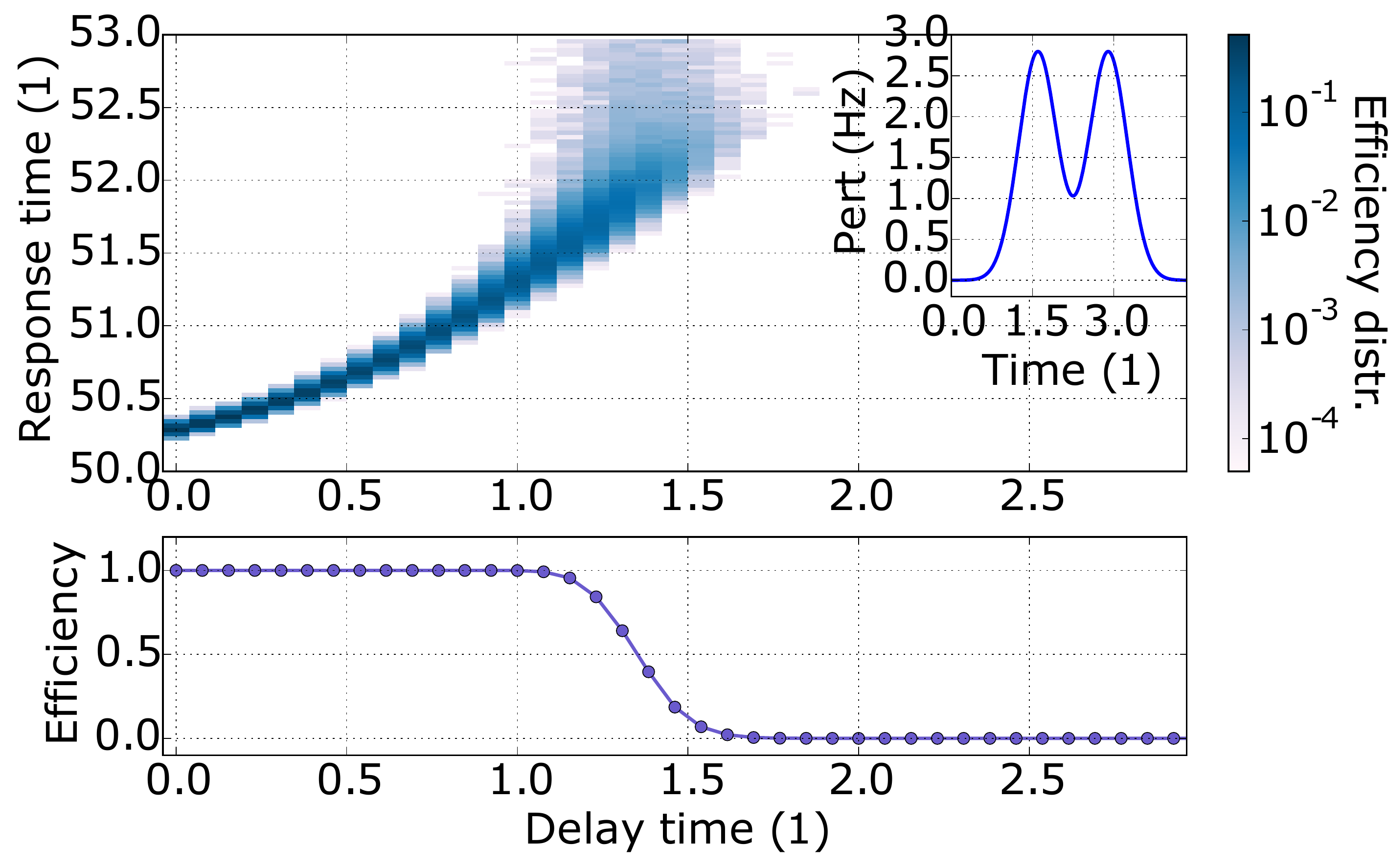}\\
\includegraphics[width=0.47\textwidth]{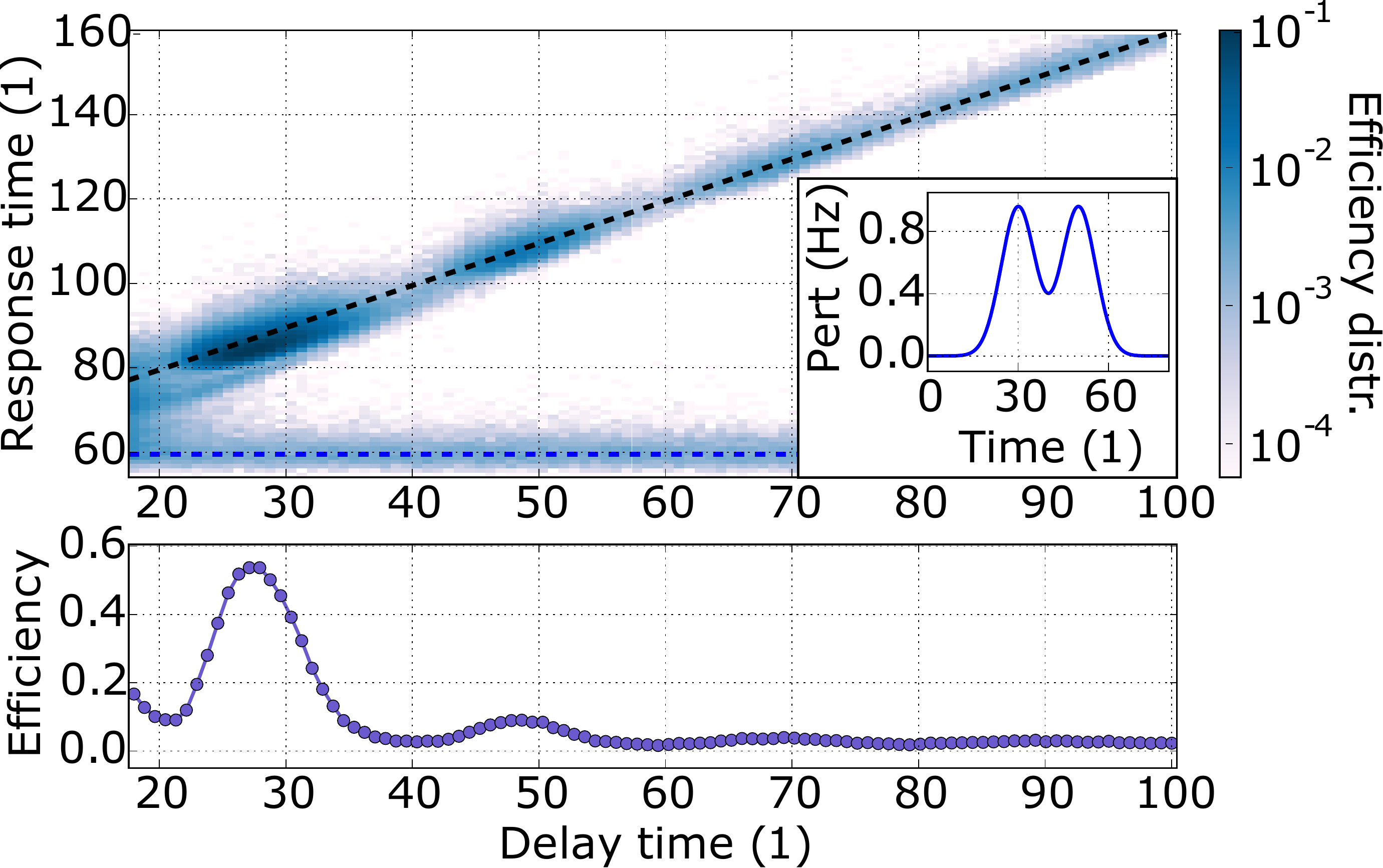}
	\caption{\label{fig_adler_hist} Integrator behaviour of the Adler equation (top) and resonator behaviour in presence of inertia (bottom). Top: Numerical response time histogram and efficiency curve of Eq.(\ref{eq_adler_sim}), with $\omega=0.01$ and $\beta=0.08$, constructed from $10\,000$ events for each delay. (Inset) Shape of the perturbation for a delay of 1.3 (at approximately 0.5 efficiency). Each of the two Gaussians has an amplitude of 2.8 Hz and standard deviation of 0.35. Bottom: Numerical response time histogram and efficiency curve curve of Eq.(\ref{eq_adler_sim_con_inerzia}), with $\omega=0.01$, $I=10$ and $\beta=0.08$, constructed from $20\,000$ events for each delay. The perturbation is made of two Gaussians with amplitude of 0.96 Hz and standard deviation of 5.65.}
\end{figure}

First we check in the Adler model  (a Class 1 neuron model, equation \ref{eq_adler_sim}) that an \textit{integrate-and-fire} response should follow a pair of perturbations. This \textit{integrator} property can be seen in the numerical simulations on Fig.~\ref{fig_adler_hist}. Here we simulate the perturbed Adler model with white noise shown in eq.\ref{eq_adler_sim}. The integration algorithm is the Euler-Maruyama method with Gaussian white noise $\langle\xi(t)\rangle=0$, ${\langle\xi(t)\xi(t-\tau)\rangle=\beta\delta(\tau)}$ with $\beta=0.08$ as the weight parameter of the random variables.

\begin{equation}
\dot{\phi}=\omega+\Delta \omega(t) - \sin\phi + \xi(t)
\label{eq_adler_sim}
\end{equation}

where $\Delta \omega(t)$ is the perturbation, with the shape of two Gaussians, witch is applied with different delays. The efficiency curve shows that, for small delays, the two perturbations are added and we observe a phase jump of $2\pi$ (an excitable event), while for delays larger than 1.5, we see no response. This means that, when the two perturbations are close enough, they get integrated and are able to overcome the threshold and produce a response.

\begin{equation}
I\ddot{\phi}+\dot{\phi}=\omega+\Delta \omega(t) - \sin\phi+\xi(t)
\label{eq_adler_sim_con_inerzia}
\end{equation}

In presence of an inertial term as in eq.~\ref{eq_adler_sim_con_inerzia}, the response of the system to pairs of perturbation changes drastically. Instead of the monotonous increase in the efficiency upon delay reduction as in the pure Adler model (Fig.~\ref{fig_adler_hist}, top panel), several maxima are easily observed for separations about 28, 50 and 70 time units, which indicates the resonator behavior. We note that clearly the efficiency also increases for shorter and shorter delays between perturbations (below 20 time units) but this increase is less related to the resonator nature of the system. In fact, as shown by the inset on Fig.~\ref{fig_adler_hist}, the two gaussian perturbations start to overlap for delays shorter than 20 time units and the resulting perturbation is not \textit{sub-threshold} anymore.

In section \ref{sec:experiment}, we have demonstrated that the semiconductor laser with injection, often described in terms of the Adler equation when discussing excitability, can present a resonator behavior. This resonant feature is absent from the pure Adler model, which is known as an integrator neuron. From the analysis above, we conclude that a small inertial term (absent in the pure Adler, overdamped limit) is sufficient to recover the resonator behavior and to recover to some extent the analogy between a mechanical and an optical system.

\section{Laser model: from integrator to resonator}
\label{sec:resonator}

Beyond the ad-hoc modelling presented above, further insight in the dynamics of the laser device can be gained by analyzing the following set of dynamical equations used in \cite{garbin2017refractory} to analyse the integrator behavior.
\begin{equation}
\begin{array}{lcl}
\dfrac{dE}{dt} &=& \sigma\left[E_I+\left(1-i\alpha\right)DE-(1+i\theta)E\right]+\xi(t)\,,\\
\vspace{-2mm} \\
\dfrac{dD}{dt} &=& \mu-\left(1+|E|^2\right)D\,,
\end{array}
\label{eq_prati}
\end{equation}
where $E$ (complex variable) is the slowly varying envelope of the electric field, $D$ (real variable) is the population variable proportional to the excess of carriers with respect to transparency. This system was also integrated using the Euler-Maruyama algorithm with $\beta=0.01$ as the noise coefficient of the system. Note that in the case of the $D$ variable, the noise was set to zero, as in was done in \cite{garbin_refractory_2017}. This is because the physically relevant noise source is the noise present in the field, and not in the population.

\begin{figure}[t] 
\includegraphics[width=0.40\textwidth]{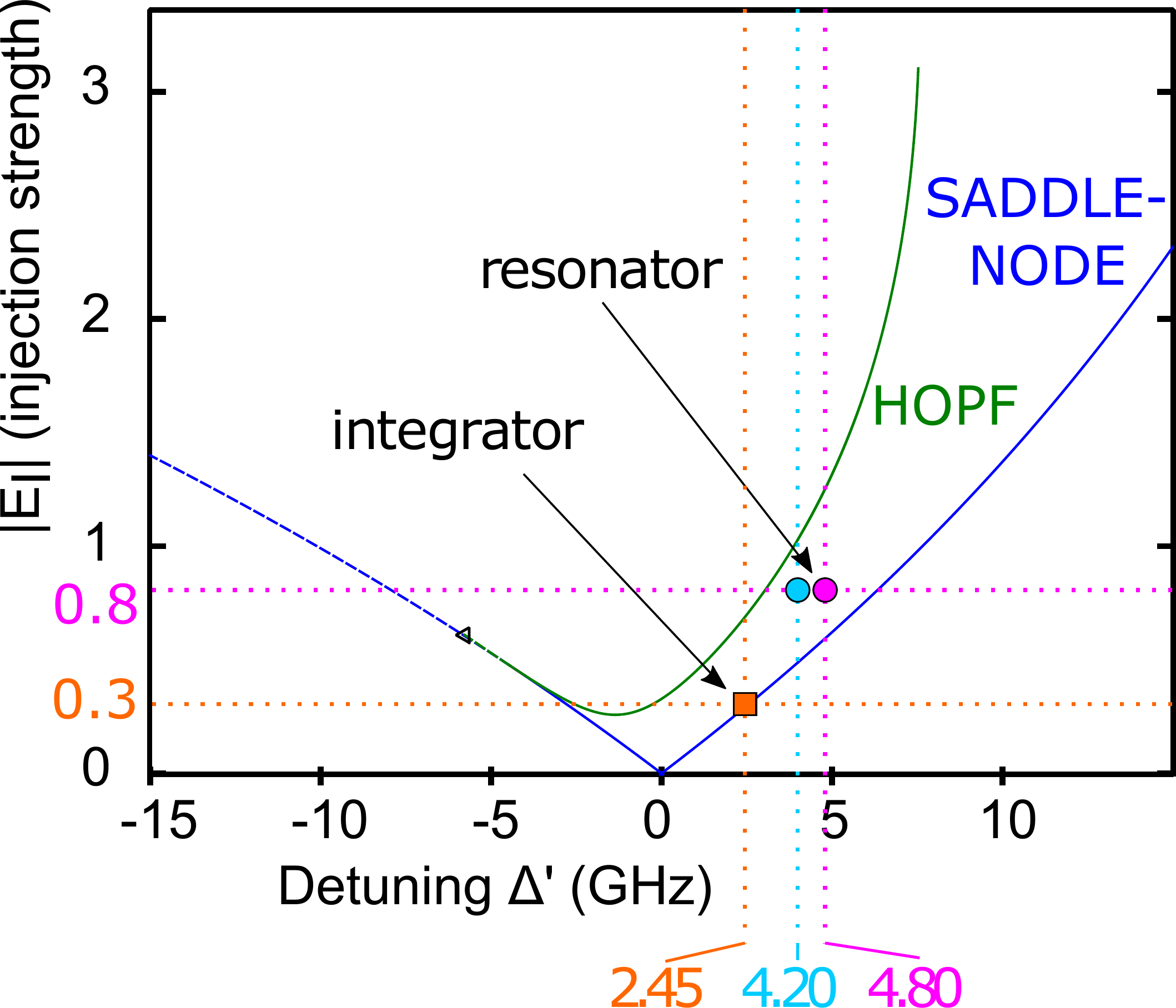}%
	\caption{\label{fig_bifurcation} Bifurcation diagram of principal codimension 1 bifurcations, with $\alpha=4$, $\sigma=50$ and $\mu=15$. The diagram shows a Fold-Hopf bifurcation, where a saddle-node and a Hopf bifurcation collide on a single point. The parameters for the simulations are chosen in the region in between the Hopf and the SN bifurcation. The pink (right) and cyan (left) points corresponds to the parameter set in which the resonator behavior is analyzed (pink being closest to the experimental observation) while the orange square shows the parameter set we chose for the integrator regime.}
\end{figure}

The physical parameters are $\alpha$, which is the linewidth enhancement factor, and $\sigma$, which is $\sigma=\tau_c/\tau_p$ where $\tau_p$ is the photon lifetime, and $\tau_c$ is the carrier lifetime. The time of the simulations is scaled to the carrier lifetime.

The three control parameters of the experiment here are denoted by $\theta$, $\mu$ and $E_I$.
$E_I$ is the dimensionless complex amplitude of the externally applied field, $\mu$ is the pump parameter of the slave laser proportional to the excess of injected current $I_{sl}$ with respect to the threshold $I_{th}$,
and the cavity detuning $\theta$ which is related to the experimental detuning $\Delta=\nu_S-\nu_M$ (defined as the frequency of the slave laser minus that of the master) by
\begin{equation}
\theta=-\alpha+2\pi\Delta\tau_p=-\alpha+\frac{2\pi\Delta'}{\sigma}\,,\qquad \Delta'=\Delta\tau_c\,.
\end{equation}
Assuming $\tau_c=1$ ns, $\Delta'$ is just the detuning in GHz.
In the simulations we fixed the physical parameters $\alpha=4$, $\sigma=50$ (\textit{i.e.} $\tau_p=20$ ps if $\tau_c=1$ ns). The optical injection strength was then set to $\mu=15$, and we chose the input intensity $|E_I|$ to be either 0.3 or 0.8, with the phase of the injected field equal to zero ($\phi_I=0$).  The detuning $\Delta'$ was chosen as to be very close to the saddle node transition in the bifurcation diagram of Fig. \ref{fig_bifurcation}, but not too far from the Hopf bifurcation. We chose two different values: $\Delta'=4.2$ and $\Delta'=4.8$, and the difference between the two cases will be discussed later.

\begin{figure}[t] 
\includegraphics[width=0.42\textwidth]{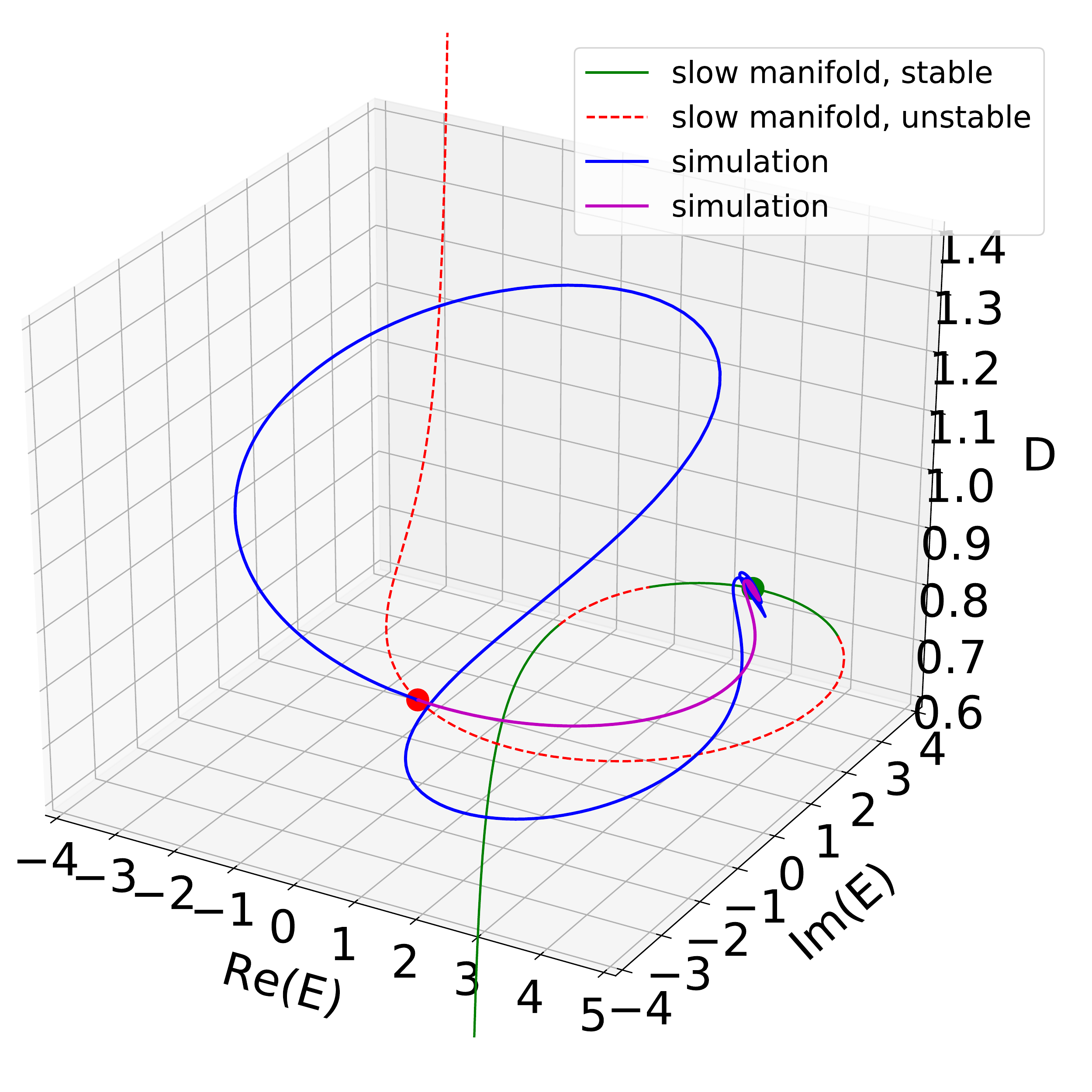}
\includegraphics[width=0.42\textwidth]{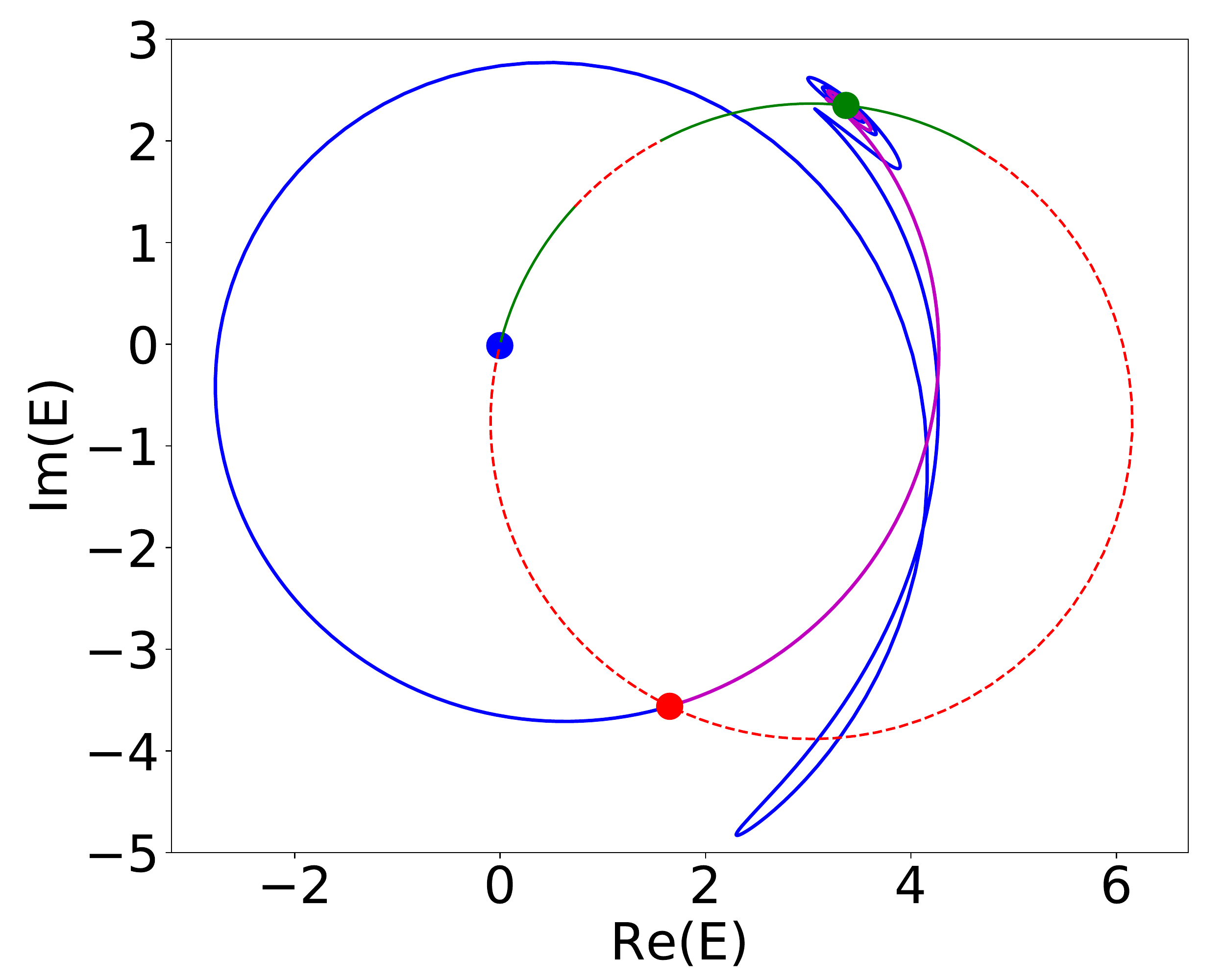}%
	\caption{\label{fig_hetero} Slow manifold and fixed points. The green (downwards) continuous part of the slow manifold is stable, the red dashed part (upwards) unstable. The red (bottom) dot is the saddle point and the green (top right) dot is stable. The blue dot (left) is the unstable focus, which cannot be seen on the upper panel as it lies much higher in the $D$ direction of phase space. Starting very close to the red saddle point, the system can relax back to the green stable fixed point following either of the two trajectories depending on the exact initial conditions (purple or blue lines, simulations without noise starting close to the red saddle point, $\Delta'=4.2$).}
\end{figure}

\begin{figure*}[t]
\includegraphics[width=0.44\textwidth]{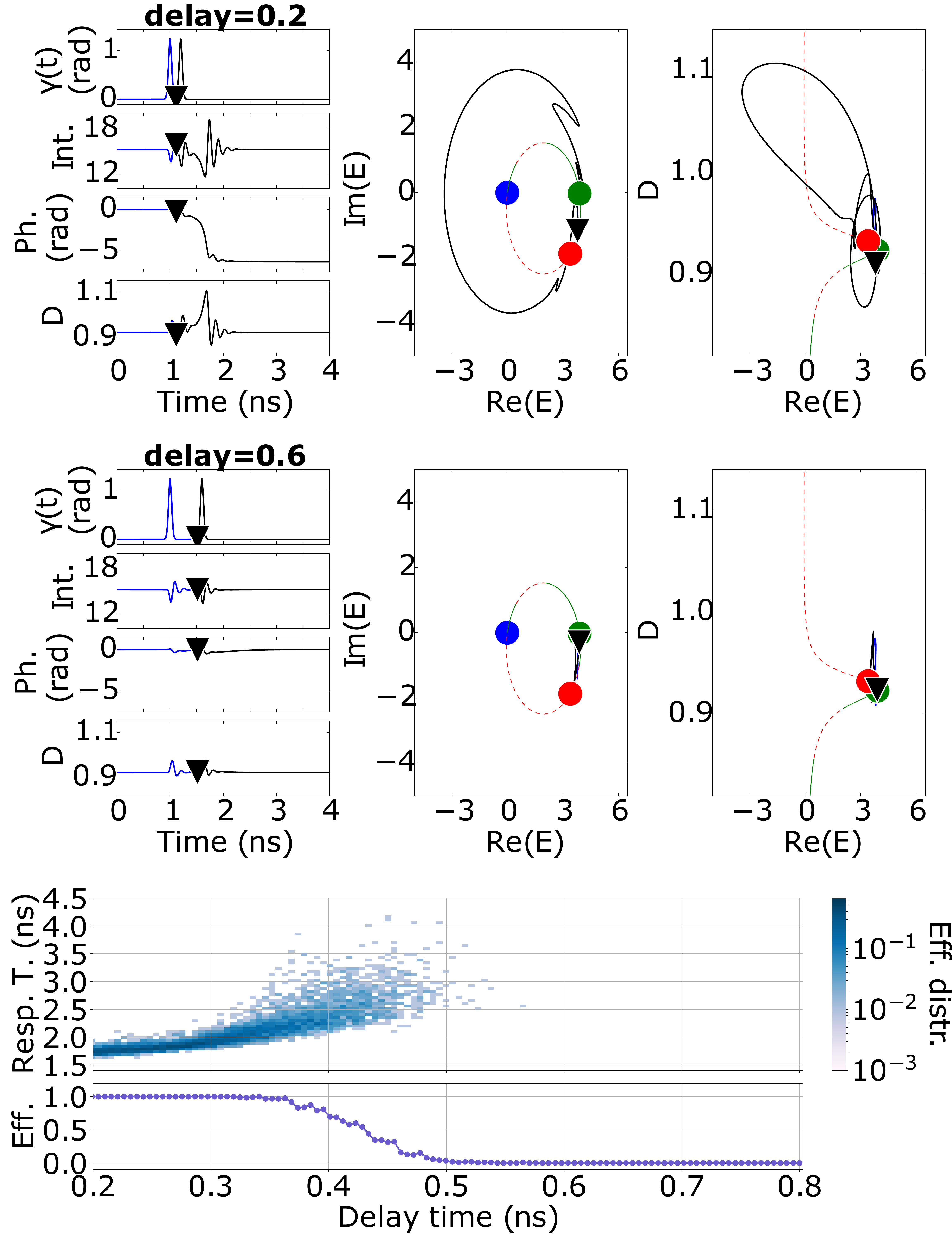}
\hfill
\includegraphics[width=0.44\textwidth]{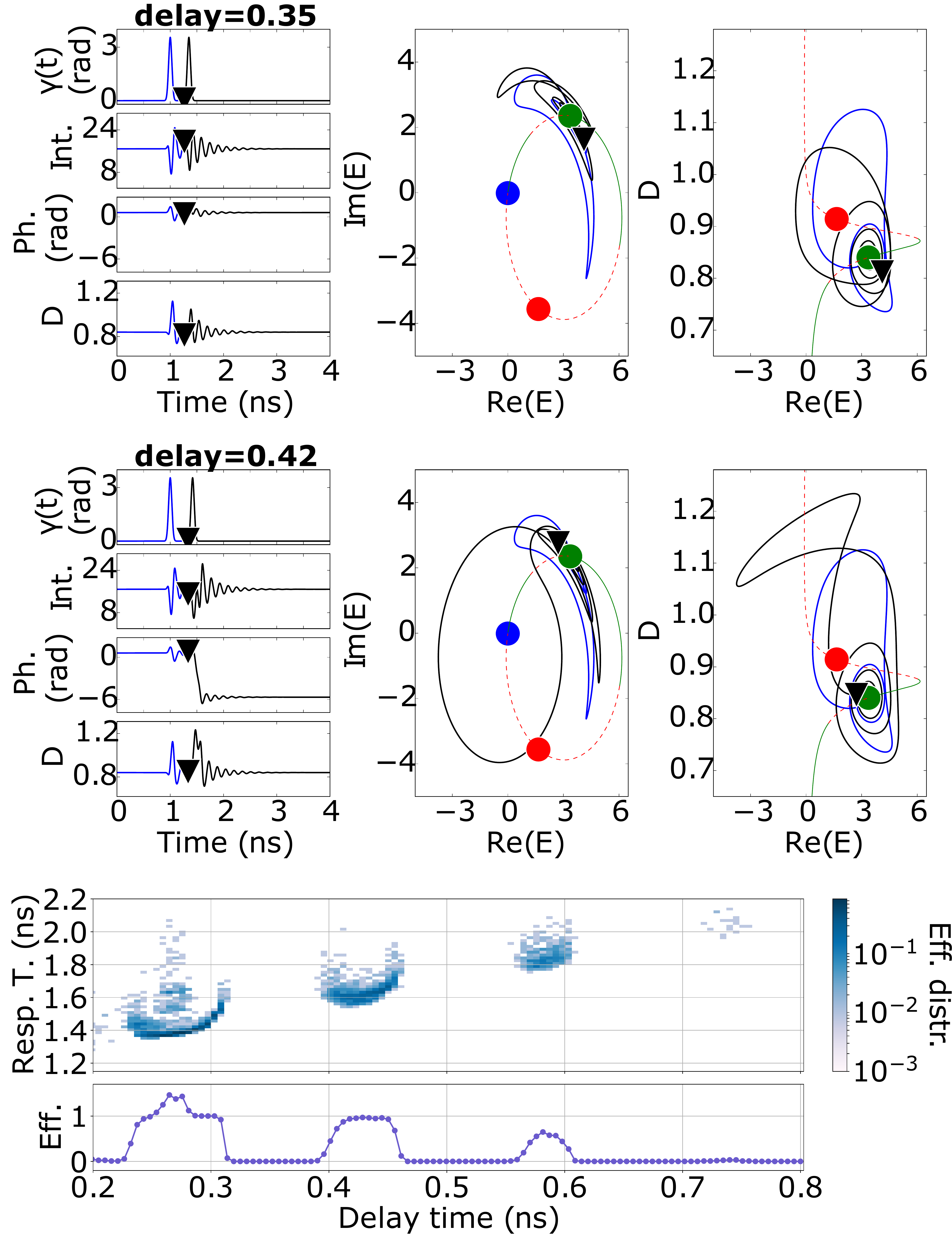}
	\caption{\label{integrator_resonator_eff} Left: Integrator example ($\Delta'=2.45$, $|E_I|=0.3$). First column: time series showing the perturbations, the emitted intensity, the relative phase and the population inversion. Second column: trajectory in the Argand plane. Third column: phase space projections on the $\Re{E},D$ plane. Black triangles indicate the occurrence of the second perturbation. Blue (left), red (bottom), green (right) dots: unstable focus, saddle, stable fixed point. Top row: two perturbations separated by 0.2~ns trigger a response (no noise). Middle row: two perturbations separated by 0.6~ns do not (no noise). Bottom row: efficiency of repeated perturbations with varying delay between them ($\beta=0.01$). Right: Resonator example ($\Delta'=4.2$). Two perturbations separated by 0.35~ns do not trigger a pulse (top, no noise), but perturbations separated by 0.42~ns do (middle row, no noise). Bottom row: the efficiency shows several maxima depending on the time separation between perturbations ($\beta=0.01$).}
\end{figure*}

As already explained in \cite{garbin_refractory_2017}, in our range of parameters the system is governed by three fixed points: un unstable focus very close to the origin in the complex plane $(\Re(E),\Im(E))$ (the blue point in FIG. \ref{fig_hetero} and \ref{integrator_resonator_eff}) and a couple of stable-unstable nodes that arise from a saddle node on invariant circle bifurcation (the green point is the stable node, and the red point is the saddle). 

Furthermore, our model can be seen as a slow-fast system. We can in fact rewrite it as:
\begin{equation}
\begin{array}{lcl}
\dfrac{dE}{dt} &=& E_I+\left(1-i\alpha\right)DE-(1+i\theta)E \equiv f(E,D)\,,\\
\vspace{-2mm} \\
\dfrac{dD}{dt} &=& \epsilon \left[ \mu-\left(1+|E|^2\right)D\ \right]\equiv g(E,D,\epsilon)
\end{array}
\label{eq_prati_slow_fast}
\end{equation}
where $\epsilon\equiv 1/\sigma=0.02$ with our choice of parameters. We then know from slow-fast systems theory and more particularly from Geometric Singular Perturbation Theory \cite{fenichel_geometric_1979,nils_berglund}, that, where the critical manifold is stable (\textit{i.e.} all the eigenvalues of the Jacobian calculated on the manifold have negative real part), the system will asymptotically converge toward the slow manifold. Here the critical manifold is defined by the parametric curve:
\begin{equation}
\begin{array}{lcl}
f(E,D)=0 &\rightarrow & E(D)=\dfrac{E_I}{(1+i\theta)-(1-i\alpha)D}\,,\\
\end{array}
\label{eq_prati_slow_fast}
\end{equation}
It has the shape of a string going from negative values of $D$ toward positive values of $D$ close to the origin, with a circular loop that develops around $0.6<D<1.0$. From a numerical analysis we know that it is stable in the green continuous regions in Fig.~\ref{fig_hetero}. Near these regions, the systems will then converge toward the slow-manifold, possibly with some oscillations. These oscillations are the ones commonly referred to as "relaxation oscillation" in laser physics, although in this specific instance they do not show the typical features of slow-fast relaxation oscillators (see~\ref{Appendix_A}). The couple of saddle-node points lie exactly on this loop.

In our simulations, we always start at the stable point. We then perturb the system with a phase-perturbation $\gamma(t)$, which has the shape of a double-Gaussian pulse, where each pulse has a variable height and a standard deviation of 0.03 ns (so that the full width at half maximum is 0.07 ns), and the delay between the two pulses is varied from 0.2 to 0.8 ns. We apply the perturbation to the phase of the injected field as follows:
\begin{equation}
\begin{array}{lcl}
\dfrac{dE}{dt} &=& \sigma\left[E_I e^{i\gamma(t)}+\left(1-i\alpha\right)DE-(1+i\theta)E\right]+\xi(t)\,,\\
\end{array}
\label{eq_prati_slow_fast}
\end{equation}
which is analogous to the phase perturbation applied in the experiment.

It is well established that in the range of parameters (especially in terms of $\Delta$ and $E_I$) close to the saddle-node bifurcation, the system can be modelled with the Adler equation, and therefore it should behave as an integrator. Further away from this parameter region, we expect the system to behave as a resonator. This transition from one to the other behavior is analysed in Fig.~\ref{integrator_resonator_eff}. The black triangle denotes the application of the second perturbation and the blue, red and green dots in phase space denote as before the unstable focus, saddle and stable fixed points respectively. In the $(\Re(E), D)$ plane, the unstable focus is not visible as it lies close to the origin in the Argand plane ($E_A\approx0$) and therefore $D_A\approx\mu=15$. On the left part of the figure, we show the integrator behavior observed for $\Delta'=2.45$ and $|E_I|=0.3$. The first two rows show two different simulations performed for a delay of 0.2, and 0.6~ns. In all cases, we start the simulation at the stable (green) fixed point, and we return at the end of the simulation time to the same point. We then apply the first perturbation and we observe that, after the effect of the perturbation, the system is displaced from the critical manifold but comes back to its original point almost without laser relaxation oscillations (better seen on the third column in a projection on the $\Re(E)-D$ plane). Here if we send a double-pulse perturbation, we find a clear integrator behavior which closely resembles the simulations of the Adler model since two perturbations which are separated by 0.2~ns trigger a response (top row) while two perturbations separated by 0.6~ns (second row) do not. Repeating the simulations varying the delay and introducing a noise of $\beta=0.01$, we get the efficiency figure at the bottom row, which is very similar to that of the Adler model (Fig.~\ref{fig_adler_hist}). 

In contrast, the resonator case can be observed with $\Delta'=4.2$. The simulations are shown on the right side of Fig.~\ref{integrator_resonator_eff}. Here the first two rows show two different simulations performed for a delay of 0.35 and 0.42 ns. This time, after the first perturbation the system relaxes back towards its stable fixed point but clearly oscillates around it a few times. Again, the role of these laser relaxation oscillations is more visible in the $\Re(E)-D$ plane (rightmost panel). Quite importantly, it is the coupling between amplitude and phase due to the linewidth enhancement factor $\alpha$ which is crucial to bring the system close to the separatrix. If we apply a second perturbation (represented by the black triangle) the system is displaced again, and if the timing is right so that the second perturbation comes when the system is already going anti-clockwise during the oscillations, then the two perturbation will sum up and trigger an excitable event. This happens for a delay of 0.42~ns but not for a delay of 0.35~ns, which indicates a non-integrator behavior. As before, performing statistical analysis in presence of noise and varying the delay between the two inputs, we can observe that the efficiency of the double perturbations oscillates with the delay between the perturbations (bottom row). This is a clear example of a resonator feature, which in this case is due to the laser converging towards its stable fixed point in an oscillatory fashion. The period of these oscillations is of about 0.16 ns, which is coherent with a theoretical calculation of the laser relaxation oscillations in this system given our parameter range (see Appendix~\ref{Appendix_A} for more details).

\begin{figure}[t]
\includegraphics[width=0.47\textwidth]{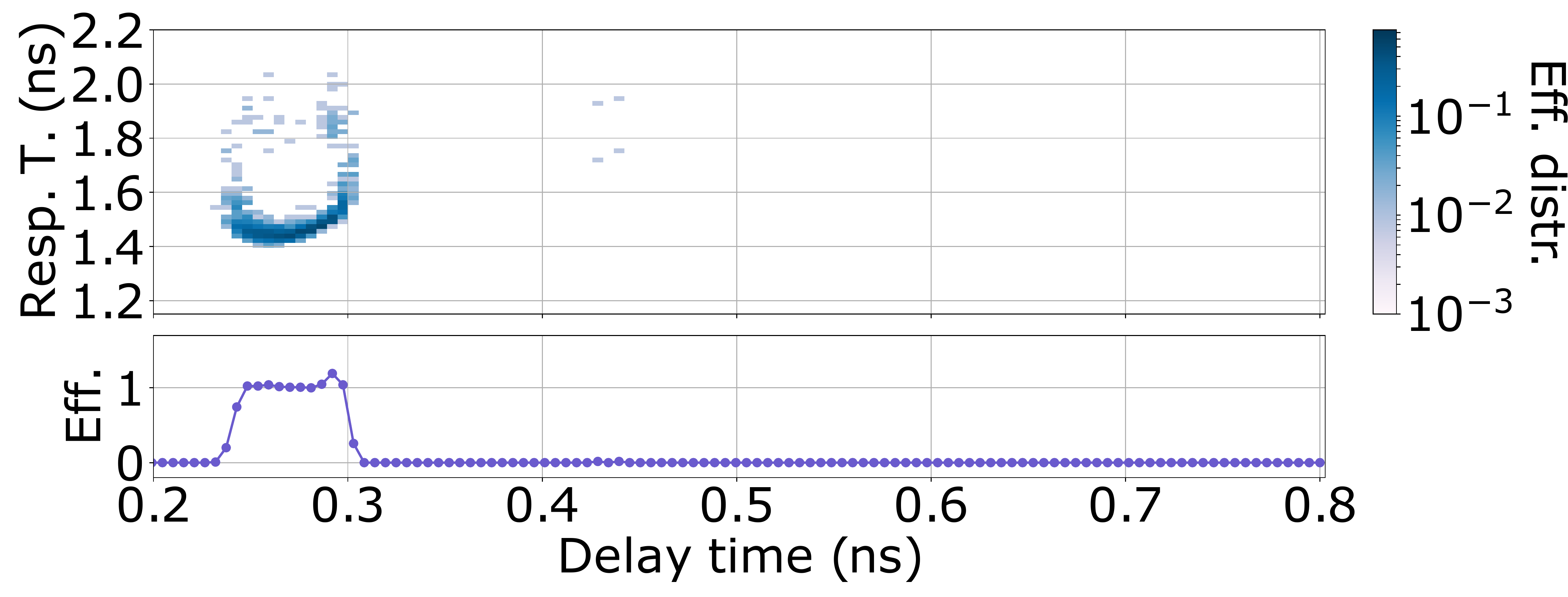}%
\caption{\label{resonance_delta_48} Resonator example and efficiency ($\Delta'=4.8$,  $\beta=0.01$)}
\end{figure}

Since the integrator behavior is clearly found when resonance features vanish, the transition between the two regimes is gradual. In fact, between these very strongly typed examples of integrator and resonator types, a simulation which closely matches the experimental findings can be obtained as shown on Fig.~\ref{resonance_delta_48}. Here the perturbation strength (\textit{i.e.} the amplitude of the pulses) is of of 2.96 radians or 169 degrees. Performing the same statistical analysis as before, we obtain a single peak in the efficiency of the perturbation. Actually, this peak is more of a plateau due to the efficiency reaching unity, with a slight bump at the end due to multipulse excitability.

As before, the origin of the bump in the efficiency curve comes from the laser relaxation oscillations, even though in the second case we do not see other bumps in efficiency for successive time delays. The reason why in the experiment we were only able to see only one single maximum is twofold. Firstly, we discarded the cases where a linear and an excitable response were not clearly separable by using an threshold in the height of the generated pulse. Secondly, the amplitude of the perturbation used in the numerics to observe a strong resonance feature is large (203 degree) compared to the second case (169 degrees), and that goes beyond the maximum amplitude that can be applied in the experiment, which is around 170-180 degrees. 

On Figs.~\ref{integrator_resonator_eff},\ref{resonance_delta_48} the efficiency is sometimes apparently larger than unity. This is due to the fact that in the simulations the detection of the excitable pulses was performed on the phase of the electric field and counting an excitable event every time there is a $2\pi$ phase rotation. Thus, the efficiency larger than unity is associated to those realizations in which the response of the system consists of more than one $2 \pi$ rotation (which we discuss later in sec.~\ref{sec:multipulse}).

\section{Separatrix}

In the previous section we have introduced the slow manifold as a reference structure that can help understand the numerical simulations. The attractive sections of the manifold are especially important, since the system will converge towards them if it is sufficiently near. Another  structure that can give us insights into the nature of the excitability of the system for different parameters is the separatrix manifold, which in this context is the 2D-surface in the 3D phase space $\Re(E)-\Im(E)-D$ that separates the regions where the system is excited from the regions where it is not. In particular, whenever the system starts from a not excited region and then crosses the separatrix, soon after it will emit one or more excitable responses. 

\begin{figure}[h!]
	
	\includegraphics[width=0.24\textwidth]{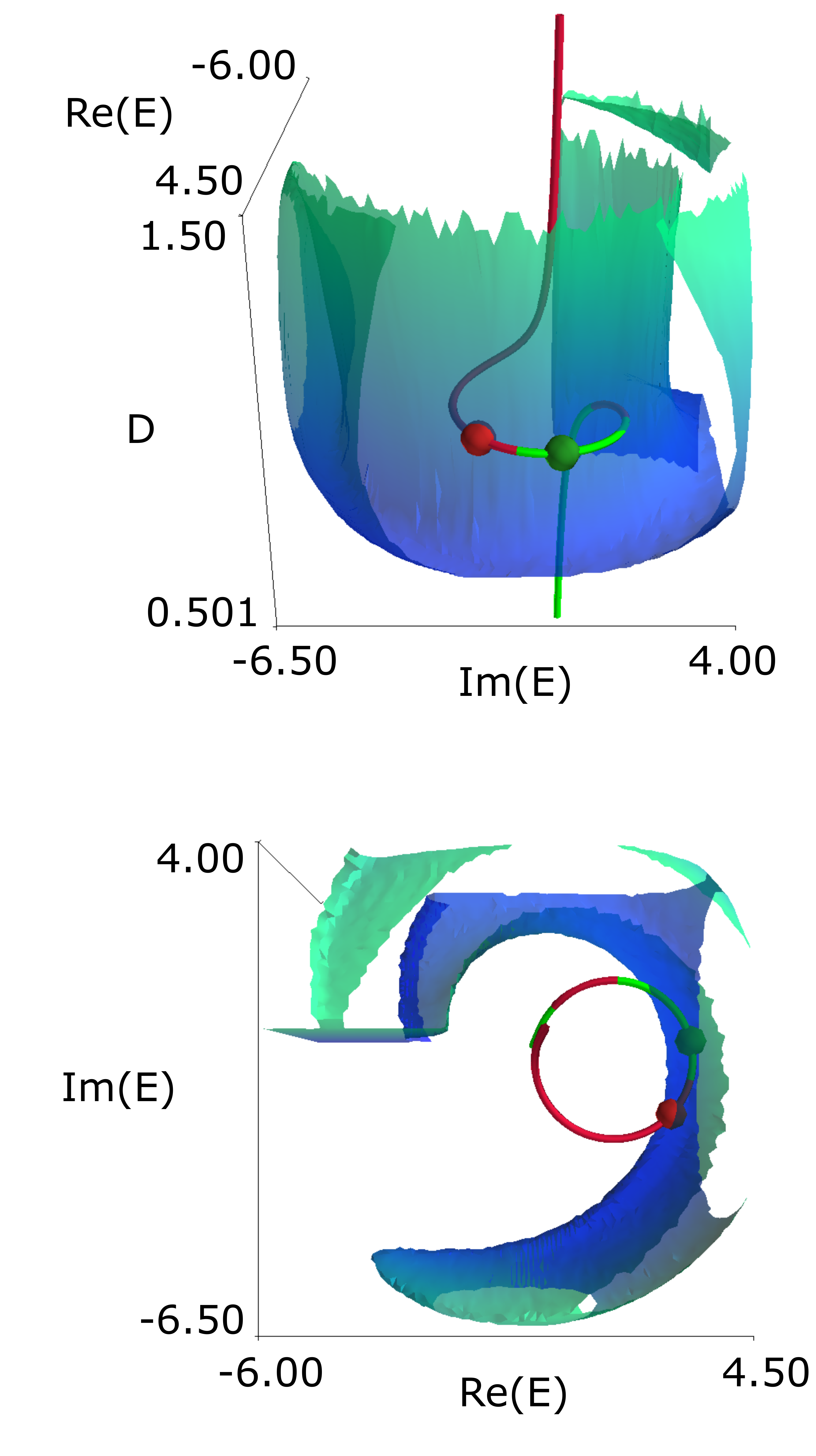}%
	\includegraphics[width=0.24\textwidth]{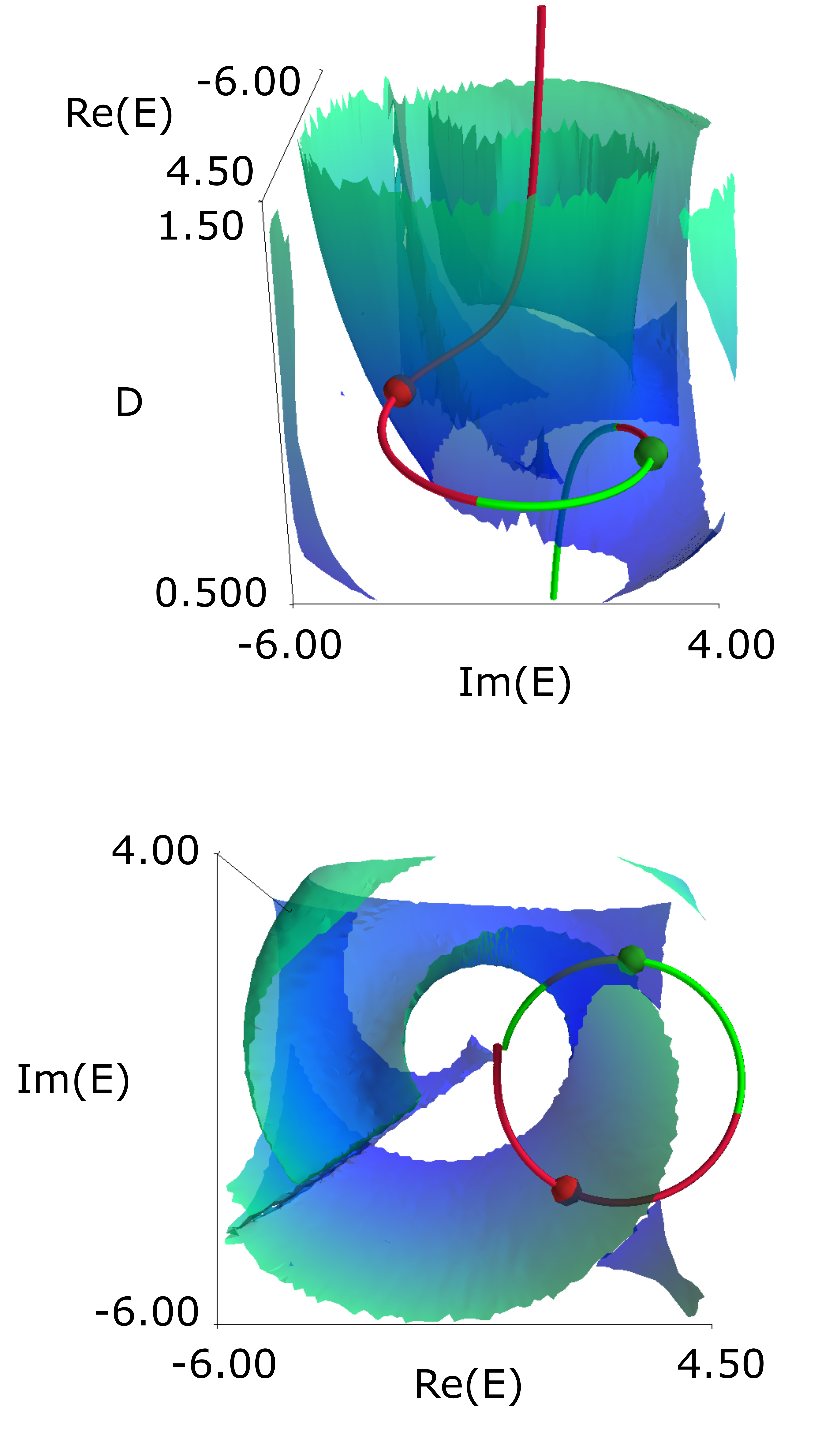}%
\caption{\label{fig_separatrix}
	3D-plot of the critical manifold, saddle-node pair and the separatrix in phase space $\Re(E)-\Im(E)-D$. Bottom figures are a top-down view of top figures. \textit{(Left figures)} Separatrix manifold in the integrator regime ($\Delta'=2.45$, $|E_I|=0.3$). \textit{(Right figures)} Separatrix manifold in the resonator regime ($\Delta'=4.2$, $|E_I|=0.8$). The color coding of the surface is proportional to $D$ to improve readability in the $(\Re(E),\Im(E))$ plane.}
\end{figure}

Fig.~\ref{fig_separatrix} displays the part of the separatrix structure that is of most interest to us, i.e. when we are close the saddle-node pair. The separatrix is calculated both for the integrator (left figures) and the resonator (right figures) set of parameters. It is computed by following the evolution of a large number of initial conditions (in this case $85^3=614125$ initial conditions) arranged in a 3D-grid. After running the numerical simulations starting from each point on the grid (without noise), we separate the ones that display at least an excitable response from the ones that don't. Each point of the grid will then be labelled with either a 1 or a 0 depending on the result, so that by the end of this procedure we obtain a 3D discrete scalar field. By employing a Marching Cubes algorithm \cite{Lorensen1987}, we can extract the polygonal mesh of the isosurface that separates the two sets, and plot it as a 2D surface. 

By looking at the shape of the separatrix with respect to the saddle-node pair, we can better interpret the behavior of the system as an integrator or a resonator. When observed from the stable node where the system initially rests, in both cases the separatrix surface has the shape of an open tube that is mostly parallel to the $D$ direction. It envelops large segments of the slow manifold close to $E=0$ and intersects the slow manifold in the red saddle point. Since the system will stay on the stable point when unperturbed and it does not travel too far from the slow manifold when perturbed, the most important part of the separatrix is the surface near the stable point, which can be approximated by a curved section of a cylinder with an axis oriented along $D$. In the case of the integrator regime (left figures) this section is very close to the fixed node, so that a perturbation in the right direction can easily push the system over the separatrix and generate an excitable response. Furthermore, the relaxation oscillations have a very small amplitude, so that, given two perturbations, they will not add more efficiently for a particular delay, at variance with the integrator behavior observed in the experiment. By comparison in the resonator regime (right figures), the separatrix near the stable point is slightly far from the point itself. After a first perturbation, the relaxation oscillations will follow, and they will occur in a plane which is almost parallel to the $D$ direction. If a second perturbation is well placed in time, it can then push the system over the separatrix and trigger a response. Because of the laser relaxation oscillations and the semiconductor linewidth enhancement factor $\alpha$, in this case the timing of the perturbation is important. This is reflected in the efficiency curve of Fig.~\ref{integrator_resonator_eff}, \ref{resonance_delta_48} that display a resonance feature.

\section{Multipulse dynamics}
\label{sec:multipulse}

In the previous section we mentioned the existence of multipulse response to perturbations. These multiple spikes have already been observed in semiconductor lasers with optical injection but in general not in response to controlled perturbations. Here we show that multiple pulses can be nucleated by perturbations and that the probability to emit one, two or more consecutive spikes is controlled by the strength of the perturbation. 

\begin{center}
\setlength{\unitlength}{1cm}
\begin{figure}[h!]
	\begin{picture}(8.6,6)
		\put(0,0){\includegraphics[width=0.47\textwidth]{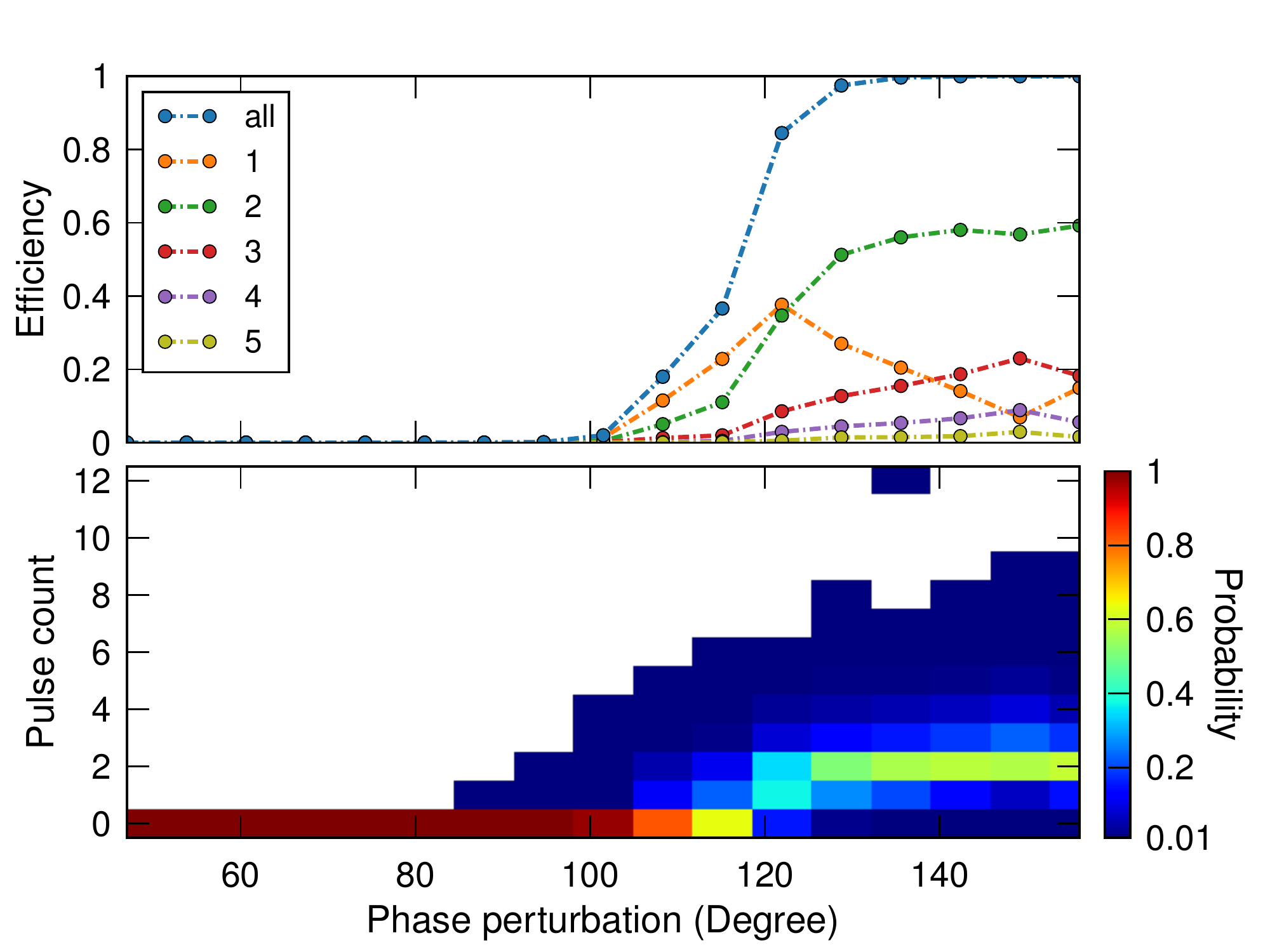}}
	\put(1.1,1.3){\includegraphics[width=0.2\textwidth]{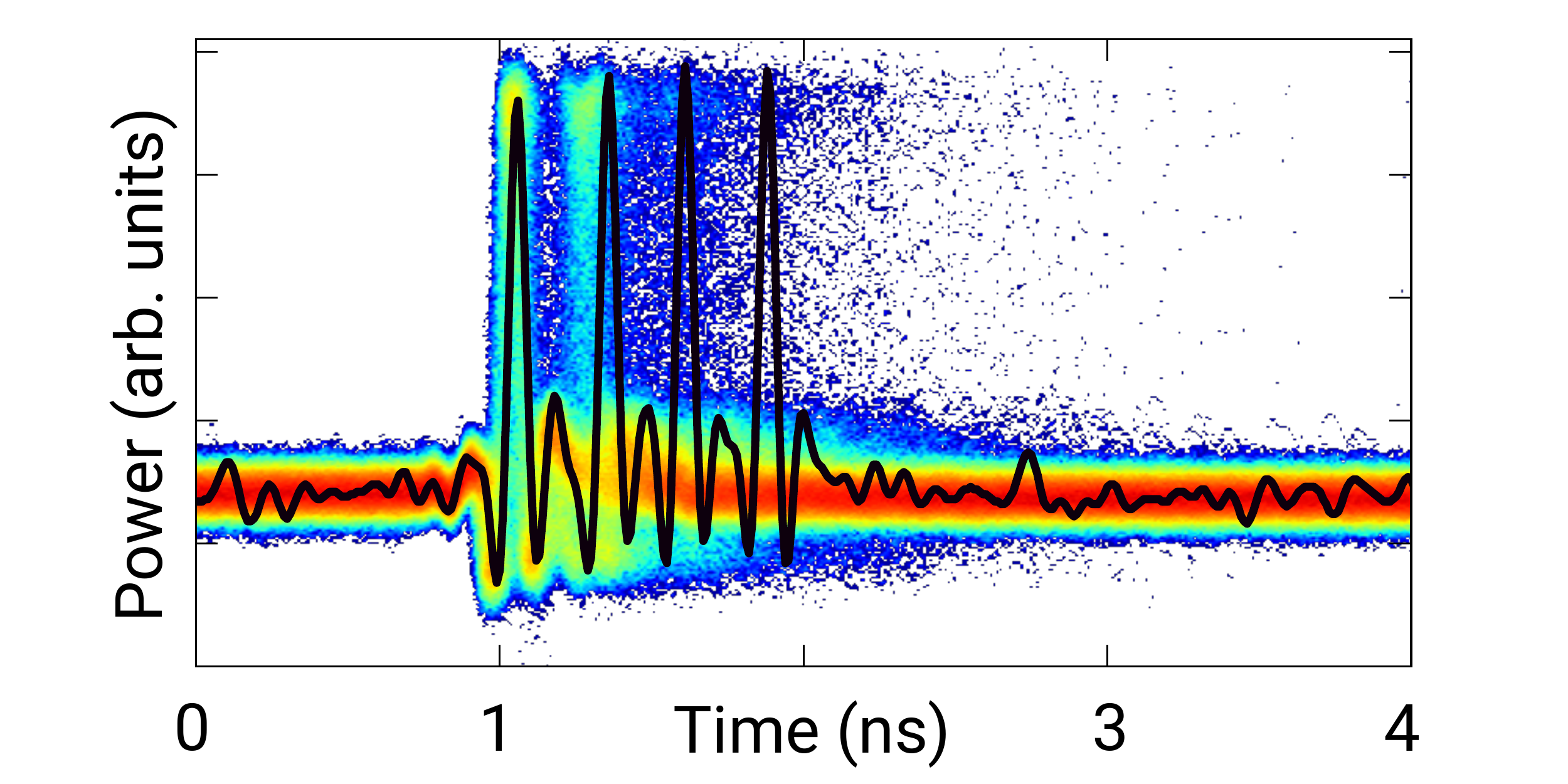}}
	\end{picture}
	\caption{Multipulse response obtained when applying strong perturbations outside of the integrator regime. Larger perturbations cause larger number of spikes, but there is a clear stochastic component in the phenomenon. Beyond 140 degrees all perturbations elicit a response but already at 120 degrees the double spike response is the most probable one. \label{fig:multi_exp}}
\end{figure}
\end{center}

We show on Fig.~\ref{fig:multi_exp} the different responses which can be obtained when moving away from the simplest excitable regime. The data was obtained by applying series of 3800 perturbations of increasing amplitude and measuring the response of the system. An example is shown on the inset where the background reflects a two-dimensional histogram of the many possible responses of the system, with an example trace as an overlay. The bottom panel shows the measured probability of 0, 1,\ldots,12 emitted pulses in response to one perturbation. For low perturbation amplitude (up to 100 degrees) no responses are detected. For increasing perturbation amplitude, single and double pulses are detected until at about 120 degrees the double pulse response become more frequent than the single pulses. The same features can be visualized on the top panel, which also includes the "total" efficiency in terms of detecting any non-zero number of spikes. It can also be appreciated that above 130 degrees, the three-spikes response becomes more frequent than the single spike, while never reaching the same value as the double spike. Of course these features can not be observed in the simple Adler model and they can be related to the carrier dynamics which is also responsible for the resonator feature. From an applicative point of view, the resonance phenomenon enables non-trivial temporal summation operation and here we see that the multipulse behavior can be used to realize an analog to digital conversion where the perturbation is converted into a series of pulses whose number is largely set by the height of the incoming pulses.

\begin{figure}[h!]
\includegraphics[width=0.47\textwidth]{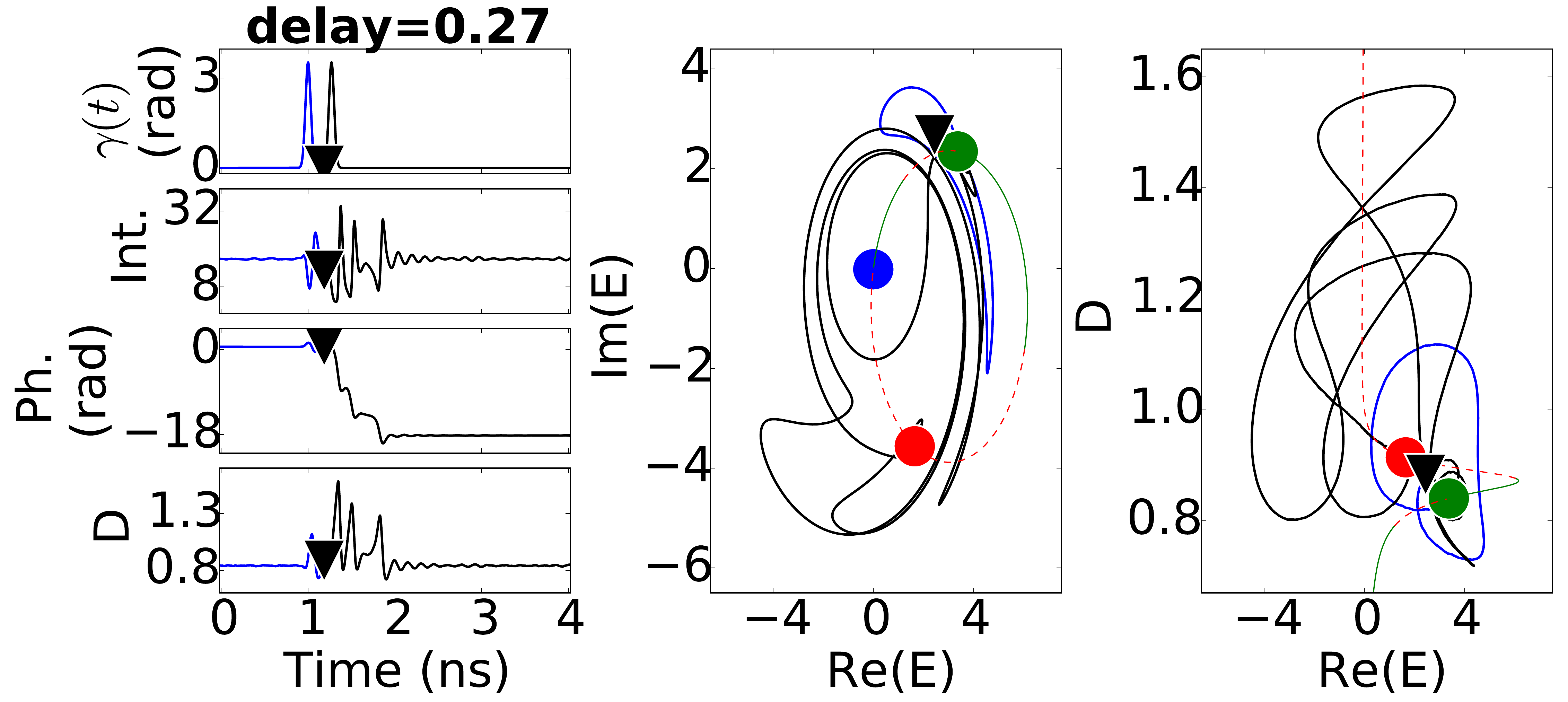}%
	\caption{\label{fig:multipulse} Multipulse response consisting of $6\pi$ phase rotation in response to perturbation ($\Delta'=4.2$, delay of 0.27~ns,  $\beta$=0.01). }
\end{figure}

Indeed, as described in section~\ref{sec:resonator}, multipulse can be observed in response to repeated perturbations in the resonator regime, as shown on Fig.~\ref{fig:multipulse}. In this case, three distinct spikes can be detected, which correspond to 3 successive rotations of the phase around the unstable branch of the slow manifold.

\section{Discussion and conclusion}

As we saw both in the experimental results and in the simulations, the system of a laser with injected signal subjected to a phase perturbation can act both as an integrator and as a resonator. This means that, if we stimulate it at a frequency which is close (or a multiple) of the relaxation oscillation period, it will respond with an excitable orbit, while for other frequencies it will not. In this regime, multipulse responses predicted in \cite{wieczorek2002multipulse} can also be observed in response to perturbations, with increasing perturbations leading to larger number of spikes in the response. However, these different responses are difficult to clearly associate to well separated homoclinic teeth. We attribute this to the effect of noise in conditions in which the homoclinic teeth may be very close to each other \cite{kelleher2011excitability}. From a laser point of view, the existence of different excitability regimes and the actual difficulty to isolate experimentally the simplest "integrate and fire" behavior of the Adler equation can be expected from the finite value of the amplitude-phase coupling $\alpha$ in quantum well media, which plays the same role \cite{zimmermann2001global} as the atomic detuning analyzed in \cite{solari1994laser}.

From a neuroscience analogy point of view, the above observations imply that the quantum-well semiconductor laser with optical injection in this range of parameters behaves more like a Class 2 neuron \cite{dynamical} than a Class 1. They are a class of neurons where the sequence of action potentials are generated in a certain frequency band that is relatively insensitive to changes in the strength of the applied current and there appears to be a minimum frequency of the generated spikes which is associated with a discontinuity in the frequency-current curve. This point also matches the experimental observation reported in \cite{bruno13} that in this experimental device the unlocking transition is in general observed at a non-strictly zero frequency. Common types of resonator neurons includes most cortical inhibitory interneurons, including the FS type, and brainstem mesencephalic V neurons and stellate neurons of the entorhinal cortex \cite{dynamical}. The models that are usually used in order to reproduce the behavior of a Class 2 neuron are those which exhibit a Hopf bifurcation, as in the case of the Fitzhugh-Nagumo model. In these types of models the existence of a discontinuity in the frequency-current curve comes from the fact that, at the bifurcation point, there is a change in dynamics from a stable point to a spiking limit cycle, which is born with a defined frequency. Following the emission of an excitable spike, such systems relax back to their stable point via oscillations which allow for a resonance effect. In biology, this type of oscillations can be observed experimentally as Membrane Potential Oscillations \cite{leung,bland}. 

Here we observe that both integrator and resonator dynamics can be obtained depending on parameters. The same type of switch from an integrator to a resonator has also been seen in neurons. In \cite{prescott_pyramidal_2008} for example, it has been observed how pyramidal neurons can switch from being integrators in vitro to resonators under in vivo-like conditions, and in \cite{zeberg_density_2015} it has been shown how a particular parameter (the density of voltage-gated potassium channels) was able to shift the dynamics of the model of the same neurons from a Class 1 to a Class 2. 

Besides this biological analogy, the integrator versus resonator properties of semiconductor lasers operated in an excitable regime may be relevant to their application in spike processing. For instance, the integrator property may be used to provide temporal summation \cite{selmi2015temporal} for phase encoded data and the resonator effect may be used to provide advanced coincidence detection feature. The capability to generate multiple pulses upon reception of larger perturbations may be used for analog to spike signal conversion, playing a complementary role of the recently demonstrated digital to spike conversion \cite{ma2017all}. Finally, these features may be relevant for the computational properties of networks built on these excitable building blocks, especially because they may strongly impact the locking dynamics of collections of excitable nodes \cite{olmi2014hysteretic}. In particular, when an excitable system is coupled to itself after a long delay \cite{garbin2015topological}, pure refractory time is expected to give rise to repulsive interactions between spikes \cite{terrien2018pulse} while resonator features may be at the origin of clusters \cite{garbin2017interactions}.

\section{Acknowledgements}
This work was conducted within the framework of the project OPTIMAL granted by the European Union by means of the Fonds Européen de développement regional, FEDER. 

The authors acknowledge support of R\'egion Provence Alpes C\^ote d'Azur through grants number DEB 15-1383 and DEB 15-1376.

\appendix

\section{About laser relaxation frequency and damping}
\label{Appendix_A}

In most settings, the term "relaxation oscillations" refers to the dynamics typically observed in the slow-fast Van Der Pol oscillator (see \textit{e.g.} \cite{ginoux2012van} for an interesting perspective and \cite{barland2003experimental} for a laser example). In laser physics the relaxation process of an unperturbed semiconductor laser towards its stable lasing solution is in general oscillatory due to the very different time scales of the electric field and carriers (see \textit{e.g.} \cite{lugiato_prati_brambilla_2015} and \cite{lippi2000invariant}). Thus the term "relaxation oscillations" is widely used even very close to the stable fixed point, where oscillations typically do not display prominently the distinctive features of slow-fast systems.

Specifically, in the case of a semiconductor laser, the small signal frequency of these oscillations can be calculated analytically \cite{lugiato_prati_brambilla_2015} as:

\begin{equation}
\Omega_{RO}=\sqrt{2\kappa \gamma_{\parallel}(a-1)}
\end{equation}
with a damping rate:
\begin{equation}
\Gamma_{RO}=\gamma_{\parallel}
\end{equation}
where $\kappa=1/\tau_p$ is the cavity damping constant (the inverse of the photon lifetime inside the cavity), $\gamma_\parallel=1/\tau_c$ is the inverse of the carriers lifetime and $a$ is the value of $D$ for the trivial stationary solution of the laser model, so that $D_s=a$ when $|E_s|^2=0$, which is the same as $\mu$ in our case.

In our simulations we assumed that $\tau_c=1$ ns, $\sigma=50$ and $\mu=15$, so that we obtain a value of the relaxation oscillations frequency:
 \begin{equation}
\Omega_{RO}=\sqrt{2\dfrac{\sigma}{\tau_c^2}(\mu-1)}= 37.42 \text{ ns}^{-1}
\end{equation}
where we have made use of the fact that $\kappa \gamma_{\parallel}=1/(\tau_c \tau_p)=\sigma/\tau_c^2$. The period of the relaxation oscillations in our case is then given by $T=\dfrac{2\pi}{\Omega_{RO}}=0.17$ ns, which is not too far from the value of 0.12 ns that was found experimentally. 

In presence of a weak injected field, the small signal of these oscillations is not altered \cite{kelleher2012modified} and only the damping rate changes, eventually leading to the Hopf bifurction. However, the frequency determined here is only valid for small linear oscillations around the stable fixed point and may not be valid for large excursions in the laser intensity or population inversion, whose period may differ markedly from that of the small amplitude oscillations \cite{lugiato_prati_brambilla_2015}.

\bibliography{biblio_resonator}%

\end{document}